\documentclass[aps,prd,twocolumn,citeautoscript,showkeys]{revtex4-1}         
\pdfoutput=1 
\usepackage{amsmath,amssymb,mathrsfs,bm,feynmf,setspace}
\usepackage{graphicx}     
\usepackage[tight]{subfigure}          
\usepackage{color} 
\usepackage{lipsum}   
\usepackage{braket}    
\usepackage[colorlinks=true]{hyperref}         
\hypersetup{
    bookmarks=true,         
    unicode=false,          
    pdftoolbar=true,        
    pdfmenubar=true,        
    pdffitwindow=false,     
    pdfstartview={FitH},    
    pdftitle={My title},    
    pdfauthor={Author},     
    pdfsubject={Subject},   
    pdfcreator={Creator},   
    pdfproducer={Producer}, 
    pdfkeywords={keyword1} {key2} {key3}, 
    pdfnewwindow=true,      
    colorlinks=true,       
    linkcolor=magenta, 
    citecolor=blue,        
    filecolor=magenta,      
    urlcolor=cyan           
} 
\definecolor{darkblue}{rgb}{0.2, 0, 0.8}
\definecolor{darkgreen}{rgb}{0.2, 0.71, 0}   
\definecolor{cadmiumred}{rgb}{0.89, 0.0, 0.13}

\numberwithin{equation}{section}

\newcommand{\req}[1]{(\ref{#1})} 
\newcommand{\labell}[1]{\label{#1}}

\newcommand{\bea}{\begin{eqnarray}}
\newcommand{\eea}{\end{eqnarray}}
\newcommand{\ba}{\begin{eqnarray}}
\newcommand{\ea}{\end{eqnarray}}

\newcommand{\beq}{\begin{equation}}
\newcommand{\eeq}{\end{equation} }  
\newcommand{\beqa}{\begin{eqnarray}}
\newcommand{\eeqa}{\end{eqnarray}}
\newcommand{\beqar}{\begin{eqnarray*}}
\newcommand{\be}{\begin{equation}}
\newcommand{\ee}{\end{equation}}
\newcommand{\eeqar}{\end{eqnarray*}}

\newcommand{\eg}{{\it e.g.,}\ }
\newcommand{\ie}{{\it i.e.,}\ }

\newcommand{\rfig}[1]{Fig.\thinspace\ref{#1}}

\DeclareMathOperator{\tr}{Tr}  

\renewcommand{\href}[2]{#2}

\begin{document}  

\title{Holographic torus entanglement and its RG flow}     
\author{Pablo Bueno$^{1,2}$ and William Witczak-Krempa$^{3,4}$}
\affiliation{\vspace{0.2cm}
$^{1}$Instituut voor Theoretische Fysica, KU Leuven, Celestijnenlaan 200D, B-3001 Leuven, Belgium\\
$^{2}$Institute for Theoretical Physics, University of Amsterdam, 1090 GL Amsterdam, The Netherlands\\
$^{3}$D\'epartement de Physique, Universit\'e de Montr\'eal, Montr\'eal (Qu\'ebec), H3C 3J7, Canada\\
$^{4}$Department of Physics, Harvard University, Cambridge MA 02138, USA 
}   
\date{\today}
\begin{abstract}     
We study the universal contributions to the entanglement entropy (EE) of 2+1d and 3+1d 
holographic conformal field theories (CFTs) on topologically non-trivial manifolds, focusing on tori. The holographic bulk corresponds to
AdS-soliton geometries.
We characterize the properties of these regulator-independent EE terms
as a function of both the size of the cylindrical entangling region, and the shape of the torus.  
In 2+1d, in the simple limit where the torus becomes a thin 1d ring, 
the EE reduces to a shape-independent constant $2\gamma$. This is twice the EE obtained by 
bipartitioning an infinite cylinder into equal halves. We study the RG flow of $\gamma$ by defining 
a renormalized EE that 1) is applicable to general QFTs, 2) resolves the failure of the area law subtraction, 
and 3) is inspired by the F-theorem. We find that the renormalized $\gamma$ decreases monotonically at small coupling
when the holographic CFT is deformed by a relevant operator for all allowed scaling dimensions.
We also discuss the question of non-uniqueness of such renormalized EEs both in 2+1d and 3+1d. 
\end{abstract}   
\maketitle 
\singlespacing 

\tableofcontents        
   
\section{Introduction}  
\label{sec:intro}  

One can obtain fresh insight into strongly interacting quantum systems by directly studying 
one of their most basic properties: their quantum entanglement. In recent times, the subject of entanglement has attracted a lot of interest in several areas of physics including: condensed matter \cite{Kitaev:2005dm,levin-wen,2005PhLA..337...22H,2008PhRvL.101a0504L,2009PhRvL.103z1601F,metlitski,swap,laflorencie}, quantum field theory (QFT) \cite{holzhey,Calabrese1,Calabrese2,Casini1,Casini2,Casini3}, 
and quantum gravity, \eg \cite{VanRaamsdonk:2009ar,VanRaamsdonk:2010pw,Bianchi:2012ev,
 Balasubramanian:2013lsa,Myers:2014jia,Czech:2014wka,Headrick:2014eia,RyuTaka1,RyuTaka3,RyuTaka4,RyuTaka2}.

A simple and useful measure of entanglement is the entanglement entropy (EE) 
associated with a spatial bipartition of a state described by a density matrix $\rho=\rho_A\otimes \rho_{\bar A}$,
where $\bar A$ is the complement of region $A$.
In the case of a pure density matrix $\rho=|\psi\rangle\langle \psi|$, 
the EE quantifies the amount of entanglement between $A$ and its complement, which in turn can yield deep physical insight.
The definition of the von Neumann EE is 
\begin{align} \label{eedef} 
  S(A) =-\tr \left( \rho_A \log \rho_A \right)\,,
\end{align}    
where $\rho_A=\tr_{\bar A}\rho$ is the partial trace density matrix obtained by integrating out the degrees of 
freedom in the complementary region $\bar{A}$. Focusing on the groundstate $|\psi\rangle$ 
of a local Hamiltonian, $\rho=|\psi\rangle \langle \psi|$ will generally have an EE 
that scales with the area of the boundary, $S(A)=B\cdot {\rm Area}(\partial A)/\epsilon^{d-2}+\dotsb$, where $d$ is the spacetime dimension,
and $\epsilon$ is a short distance regulator.
A $\log({\rm Area}(\partial A))$ enhancement appears when a finite fermion density is present. 
The general idea is to study the subleading terms, which in many cases contain well-defined, \ie regulator-independent, information. 
A particularly relevant instance corresponds to the EE of a disk in a CFT defined in infinite Minkowski space $\mathbb{R}^{1,2}$. 
In that case, the EE contains a subleading term, $-F$, which is both independent of $\epsilon$ and the disk's radius.
It further coincides with the free energy of the same theory on $S^3$ \cite{Casini:2011kv}. As proven by Casini and Huerta \cite{CH_F} --- see also \cite{Casini:2015woa} --- a renormalized version of this quantity \cite{Liu2012} 
is monotonously decreasing under the \emph{entire} RG flow connecting two fixed points,  
and it coincides with $F$ at fixed points. This ``F-theorem'' is one of the most celebrated applications of EE to QFT, and generalizes the earlier EE-based proof of the two-dimensional ``c-theorem'' \cite{Casini:2004bw,Casini:2016udt}. Extensions of these monotonicity theorems to CFTs defined on $\mathbb{R}^{1,d\geq 3}$ relying on the EE of smooth surfaces --- typically spheres ---  have been also proposed, see \eg \cite{Cardy:1988cwa,Solodukhin:2008dh,sinha10,sinha11}. 

Note however that smooth curved surfaces are not ideally suited for finite-size numerical calculations. 
For example, the pixelization of a disk in $d=3$ leads to corners that obscure the constant part of the EE as they contribute $\log(L/\epsilon)$
terms (in a scale-invariant theory), see \eg \cite{Fradkin:2006mb,Casini3,Casini4,Hirata,Myers:2012vs,Bueno1,Bueno2}. 
Further, one needs to ensure that the entire space is sufficiently large compared with the  
region $A$ under study, otherwise finite-size effects will alter $S(A)$.
A natural alternative is to work with entirely flat but finite entangling surfaces, which can be realized on topologically non-trivial spaces. 
In this paper we shall work with space being compactified into a torus, $\mathbb T^{d-1}=L_x\times L_1\times \cdots \times L_{d-2}$, and we take region $A$ to be 
a cylinder wrapping 
$(d-2)$ cycles: $A = L_A\times \mathbb T^{d-2}$, where $L_A$ is the length of the cylinder. Region $A$ will thus have two flat 
and compact entangling surfaces. 
We illustrate the geometry in \rfig{fig:torus3d} for $d=3$. The EE corresponding to this geometry, computed for the vacuum state 
of a theory at its conformal fixed point, is
\begin{align}  \label{gen-torus}
  S(A) = B\, \frac{{\rm Area}(\partial A)}{\epsilon^{d-2}} - \chi(\theta;b_i) +\dotsb 
\end{align}
The subleading term, $-\chi$, is the regulator-independent (universal) contribution that will be at the center of our discussion. It is 
a dimensionless function that depends on $(d-1)$ aspect ratios:
$\theta=2\pi L_A/L_x$ is the angle that specifies how much of the cycle $A$ covers, \rfig{fig:torus3d}, and the $b_i=L_x/L_i$ are
the $(d-2)$ aspect ratios that characterize the shape of the entire torus, $\mathbb T^{d-1}$. 
Different CFTs will have different functions
$\chi(\theta;b_i)$, which means that $\chi$ can be used as a highly non-trivial fingerprint of the state.
The structure in \req{gen-torus} arises because each of the two boundaries of $A$ is a flat, smooth, and compact manifold.

For fixed aspect ratios, one can ask how $\chi$ behaves under renormalization group (RG) flow. 
Indeed, it has a non-trivial potential to count degrees of freedom since on one hand it is finite for 
a large class of topological quantum field theories (TQFT), such as Chern-Simons theories in $3d$ \cite{Fradkin08,grover2011entanglement}.    
On the other hand, it is also finite at massless conformal fixed points \cite{metlitski,Chen14,will-torus,WWS16}.     

\begin{figure}
  	\centering
  	\includegraphics[scale=0.29]{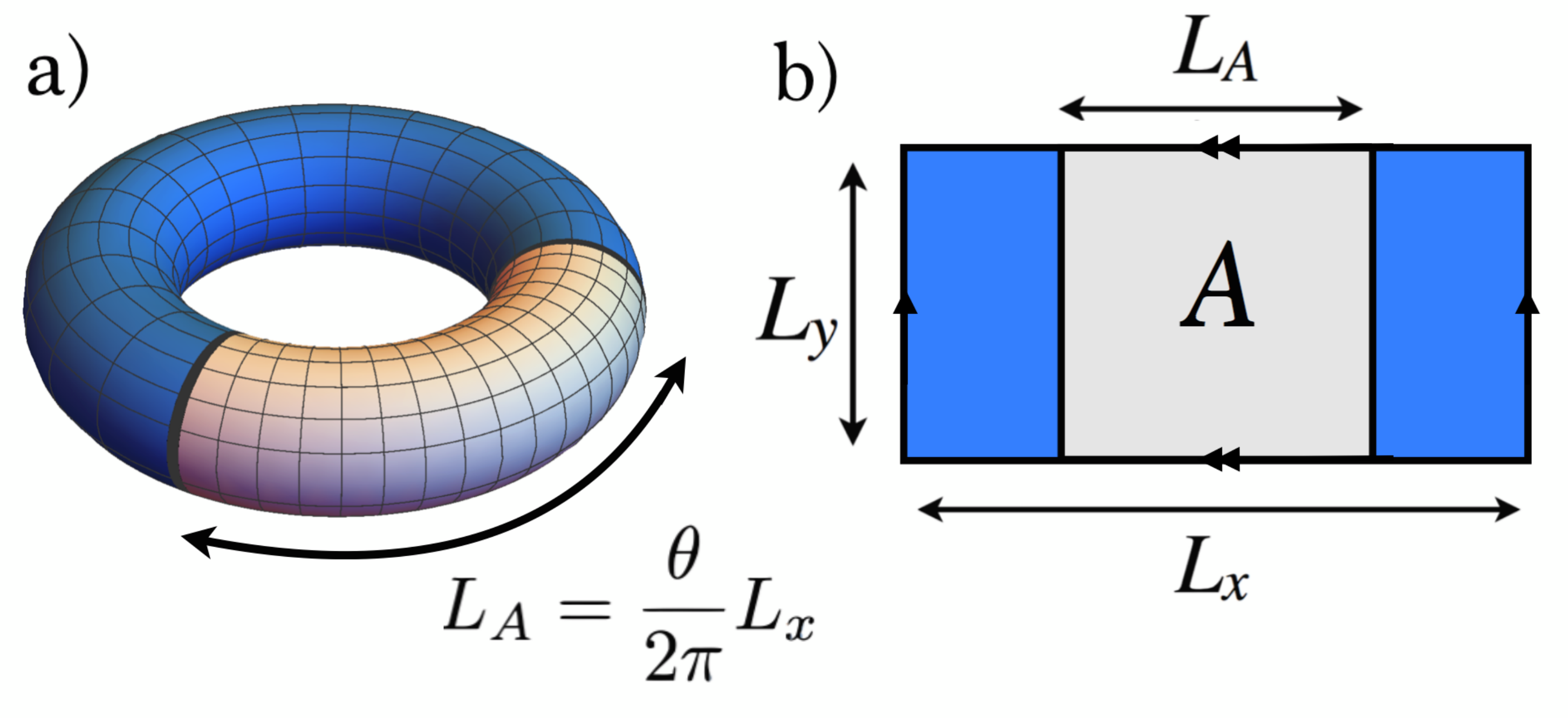}   
  	\caption{a) In a $d=3$ spacetime, we compactify space into a 2-torus. The entangling region $A$
  		is the cylinder extending along the $x$-direction. b) Since the spatial manifold is flat, a more faithful depiction is
  		on the plane, with opposite sides identified.} 
  	\labell{fig:torus3d}
\end{figure}
  
We shall mainly study $\chi$ in holographic theories. The EE of CFTs dual to Einstein gravity in the bulk 
and defined in the boundary of locally-asymptotically AdS$_{d+1}$ spacetimes can be obtained using  
the extremal-area  prescription of Ryu and Takayanagi \cite{RyuTaka1,RyuTaka2}:
\be 
S(A)=\underset{\mathcal M\sim A}{\text{ext}} \frac{{\rm Area}(\mathcal M)}{4G} \, ,\labell{RyuTaka}
\ee
where $\mathcal M$ is a codimension-$2$ bulk surface homologous to $A$ in the boundary 
(in particular $\partial \mathcal M = \partial A$). 
We will study the EE in the vacuum state, for which the relevant geometries with $\mathbb{T}^{d-1}$ topology in the spatial dimensions are the so-called AdS solitons --- see \req{solitoni}.  
 

{\bf Outline \& summary:} In section \ref{adsoli} we give general properties of the AdS-soliton geometries.  
Then, in section \ref{sec:t3} we study the properties of $\chi$ as a function of the angle $\theta$ and the aspect ratio $b$ for $3d$ 
holographic theories ($4d$ bulk). 
The cases $b\leq 1$ and $b \geq 1$ are treated in subsections \ref{bl1} and \ref{bg11} respectively, while we study the thin torus
limit $b\to \infty$ in section \ref{cyl3}. The analogous analysis is performed for $4d$ CFTs in section \ref{sec:torus4}. 

In section \ref{reer} we define new families of \emph{renormalized} EE in $3d$ and $4d$, 
which allow one to study the RG flow of the torus entanglement quantity $\chi$. This can be applied to general QFTs, not only holographic ones.  
In section \ref{holorg}, we apply this renormalization prescription in $3d$ by considering the simple
limit where the torus becomes a thin one-dimensional ring. 
The EE $\chi$ reduces to a shape-independent constant $2\gamma$; twice the EE obtained by
bipartitioning an infinite cylinder in two. 
We then deform our holographic CFT with a scalar perturbation and compute the change of the REE along the RG flow. 
We show that at leading order in the deformation, $\gamma$ decreases monotonically from its fixed point value. In section \ref{estimate} we present
an intuitive estimate of $\gamma$ in $3d/4d$ using the \emph{thermal} entropy of the infinite cylinder, and find 
that it is very close to the exact value. Appendix \ref{app1} contains a derivation of the renormalized entanglement entropy 
for a disk in a three-dimensional holographic theory deformed by a relevant operator, which we contrast with our more involved torus calculation.
Finally, in Appendix~\ref{higherdim} we study the  properties of the holographic $\chi$ in general dimensions.     

\section{Basic properties of entanglement entropy on tori} \label{sec:basics}
We review basic properties of the EE associated with a 2-cylinder bipartition of a torus in general dimensions. 
This discussion applies to all CFTs, not only those with a classical holographic description. 
Given the groundstate of a CFT in $d>2$ defined on the torus $\mathbb T^{d-1}$, the EE takes the form \req{gen-torus}, where $-\chi(\theta; {b_i})$ 
is the universal, \ie cutoff-independent contribution. By purity of the groundstate, this function is symmetric 
under reflection about $\pi$, $\chi(2\pi-\theta)=\chi(\theta)$. 
The strong subadditivity of the EE implies that $\chi(\theta)$ is convex decreasing on 
for $0<\theta\leq \pi$ \cite{will-torus}: 
\begin{align}  \label{convex}
  \chi'(\theta) \leq 0, \qquad \chi''(\theta)\geq 0 \, .
\end{align}

We now discuss the limits where region $A$ is either of maximal length ($\theta=\pi$), or shrinks to zero ($\theta\to 0$).
In the former case, the $\theta\to \pi$ limit is non-singular, which when combined with the reflection symmetry about
$\theta=\pi$, leads to the following expansion \cite{will-torus}
\begin{align}
  \chi(\theta) &= \sum_{\ell=0} c_\ell\cdot (\theta-\pi)^{2\ell}\, , \label{smooth-lim} \\
  c_0 &= \chi(\pi)\,,\nonumber
\end{align}
where only even powers appear; we have omitted the dependence on the $b_i=L_x/L_i$. By virtue of \req{convex}, 
the $c_1$ coefficient must be non-negative for all values of $b_i$. We note that nothing constrains the sign of $\chi(\pi)$,
and it can be either positive, negative or zero.
In the so-called ``thin-slice'' limit, $\theta\to 0$, we instead get the divergence \cite{will-torus} 
\begin{align}  \label{thin-slice}
  \chi(\theta\to 0) = \frac{(2\pi)^{d-2}\, \kappa}{\theta^{d-2} \, b_1\cdots b_{d-2}}\, ,
\end{align}
which is independent of the boundary conditions along the cycles. This limit produces the universal coefficient $\kappa$,
which also arises for different entangling geometries.
Indeed, in this limit the EE reduces to the EE
of a thin slab of width $L_A=\theta L_x/(2\pi)$ embedded in $\mathbb R^{d-1}$. 
The slab has infinite area, but the entropy per unit 
area is finite,  $S_{\rm slab} = 2B /\epsilon^{d-2}- \kappa / L_A^{d-2}$. (In $d=3$, the slab is a thin strip in $\mathbb R^2$.)  

A particularly important limit is the thin-torus one, where $L_{i\geq 1}\to 0$ at fixed $L_A$ and $L_x$, 
which means $b_{i\geq 1} =L_x/L_i\to \infty$. 
The system thus effectively reduces to a one-dimensional ring. In the absence of zero energy modes (which is the generic case), the short
dimensions lead to a large ``compactification'' gap, and $\chi$ reduces to a $(L_A,L_x)$-independent quantity:
\begin{align}
  \chi = 2\gamma(L_2/L_1,\dots, L_{d-2}/L_1)\,, 
\end{align}
where the factor of 2 arises from the two boundaries of $A$. Since $\gamma$ is dimensionless it only depends 
on the $(d-3)$ aspect ratios characterizing each component of the boundary
$\partial A$. In $d=3$, $\gamma$ is thus a pure constant independent of any length scale, just like the 
$F$ term for the disk EE. 

Another limit discussed in \cite{W2} is the wide torus, where $L_x\to 0$, which means $b_{i\geq 1}=L_x/L_i\to 0$. 
Alternatively, we can think of this limit as arising from diverging $L_i$ at fixed $L_A$ and $L_x$.
The EE is thus expected to be dominated by the diverging length scales, $\chi \sim L_1 \cdots L_{d-2}$, or more precisely \cite{W2}
\begin{align}  \label{wide-torus}
  \chi(b_i\to 0) = \frac{f(\theta)}{b_{1} \cdots b_{d-2}}  \, ,
\end{align} 
where $f(\theta)$ is a function that depends only on $\theta$, and was made dimensionless by factoring out $1/L_x^{d-2}$.  
We note that for the above equation to be consistent with the thin slice limit \req{thin-slice}, 
we have $f(\theta\to 0)=(2\pi)^{d-2}\kappa/\theta^{d-2}$. In addition, this leads to the following small-$b_i$ scaling 
for the smooth limit coefficients defined in \req{smooth-lim}, $c_\ell= \bar c_\ell /(b_1\cdots b_{d-2})$, where the $\bar c_\ell$ are pure
constants independent of the geometry. It would be interesting to understand their relation to CFT data.        

\subsection{Relation to corner entanglement}
In the $d=3$, when working in infinite space $\mathbb R^2$, and with a region $A$ containing a corner (kink) with opening angle $\theta$, 
the EE acquires a logarithmic correction:
\begin{align} \label{corner}
  S=B \frac{L}{\epsilon} - a(\theta) \log(L/\epsilon) +\dotsb
\end{align}
where the corner coefficient $a(\theta)$ is regulator-independent and contains non-trivial information about the groundstate \cite{Fradkin:2006mb,Casini3,Casini4,Hirata,Myers:2012vs,Bueno1,Bueno2}. $a(\theta)$ shares many connections with $\chi(\theta)$ \cite{will-torus}. First, it is symmetric about $\pi$, $a(2\pi-\theta)=a(\theta)$, and decreasing convex on $(0,\pi]$, \req{convex}. 
Further, $a(\theta)$ behaves the
same way in the $\theta\to \pi$ limit: it obeys Eq.~\req{smooth-lim}, but with $a(\pi)=0$. In the sharp limit, it also has a $1/\theta$ 
divergence: $a(\theta\to 0)=\kappa/\theta$, which is the analogue of Eq.~\req{thin-slice}. We emphasize the fact that it is the same
constant $\kappa$ that appears in both divergences because the small-angle corner coefficient is also 
directly related to the EE of a thin strip \cite{Myers:2012vs,Bueno2}. It would be interesting to see if other connections exist between $a(\theta)$
and $\chi(\theta)$, such as some of the bounds that apply to $a(\theta)$ \cite{corner-bounds}.  

In $d>3$, one can also make connections between $\chi$ and higher dimensional analogues of $a(\theta)$, such as the coefficient
associated with cones in $4d$, however we shall not describe them here.

\section{Holographic torus entanglement} \label{holotorus} 
In this section we study the properties of the torus EE $\chi$  
in three and four-dimensional CFTs with holographic duals as a function of both the size of the cylindrical entangling region, 
and the shape of the torus. We study in detail the different regimes, and provide analytic expressions in certain cases. 
We start with a quick review of AdS$_{d+1}$ solitons, which are the bulk geometries relevant for our calculations.

\subsection{AdS solitons} \label{adsoli}
AdS solitons are locally asymptotically AdS$_{(d+1)}$ solutions to Einstein's equations with negative cosmological constant with a non-trivial topology in --- at least --- one of their spatial directions, which is compactified on a circle. One can also consider the additional transverse directions 
to be periodic, so the conformal boundary is foliated 
by spacelike tori $\mathbb T^{d-1}$.
These geometries can be obtained by a double-Wick rotation of the usual planar Schwarzschild-AdS 
metric (also known as AdS black brane), $t\to i x$ and $x \to i t$, where $x$ is one of the transverse spatial dimensions \cite{Soli2,Soli1}, so their metric in Poincar\'e coordinates reads
\begin{align}\label{solitoni}
  ds^2 = \frac{1}{z^2}\left[\frac{dz^2}{f(z)}+f(z)\,dx^2+d\vec{y}_{(d-2)}^{\,2}-dt^2 \right],    
\end{align}
where $f(z)=1-(z/z_h)^{d}$, $z$ is the holographic coordinate and $y_{i}$ are the remaining $(d-2)$ transverse spatial directions. 
Regularity in the bulk --- which in the $(x,z)$ plane looks like a cigar ending at $z=z_h$ --- 
fixes the $x$ period in terms of $z_h$ as $L_{x}=4\pi z_h/d$. The notation $z_h$ alludes to the connection with the
horizon of the black brane; we shall refer to $z_h$ as the pinching point. The background \req{solitoni} dominates the 
partition function only when $x$ corresponds to the compact direction with the smallest length. Otherwise, the relevant solution will be the one in which $f(z)$ appears in $g_{y_j y_j}$ when $L_j$ is the smallest length, and so on.

In the context of the AdS/CFT correspondence, these solutions have been interpreted as dual to states of Yang-Mills theories in 
which anti-periodic boundary conditions have been imposed on the fermions around the $x$ direction, thus breaking supersymmetry and giving rise to a mass gap and confinement \cite{Soli2}. In \cite{Soli1}, it was shown that AdS solitons have negative energy, as expected for the Casimir energy of the putative dual gapped theory. Further, it was conjectured that in fact they are the lowest energy solutions sharing their boundary conditions \footnote{As argued in \cite{Shaghoulian:2015lcn}, assuming that the
 AdS$_{(d+1)}$ black brane provides the holographic thermal entropy on a torus, modular invariance implies that the AdS$_{(d+1)}$ soliton is the correct ground state geometry.
}. Strong evidence in favor of this conjecture was recently provided in \cite{Galloway:2001uv}. 

\begin{figure}[h] 
  \centering
  \includegraphics[scale=0.32]{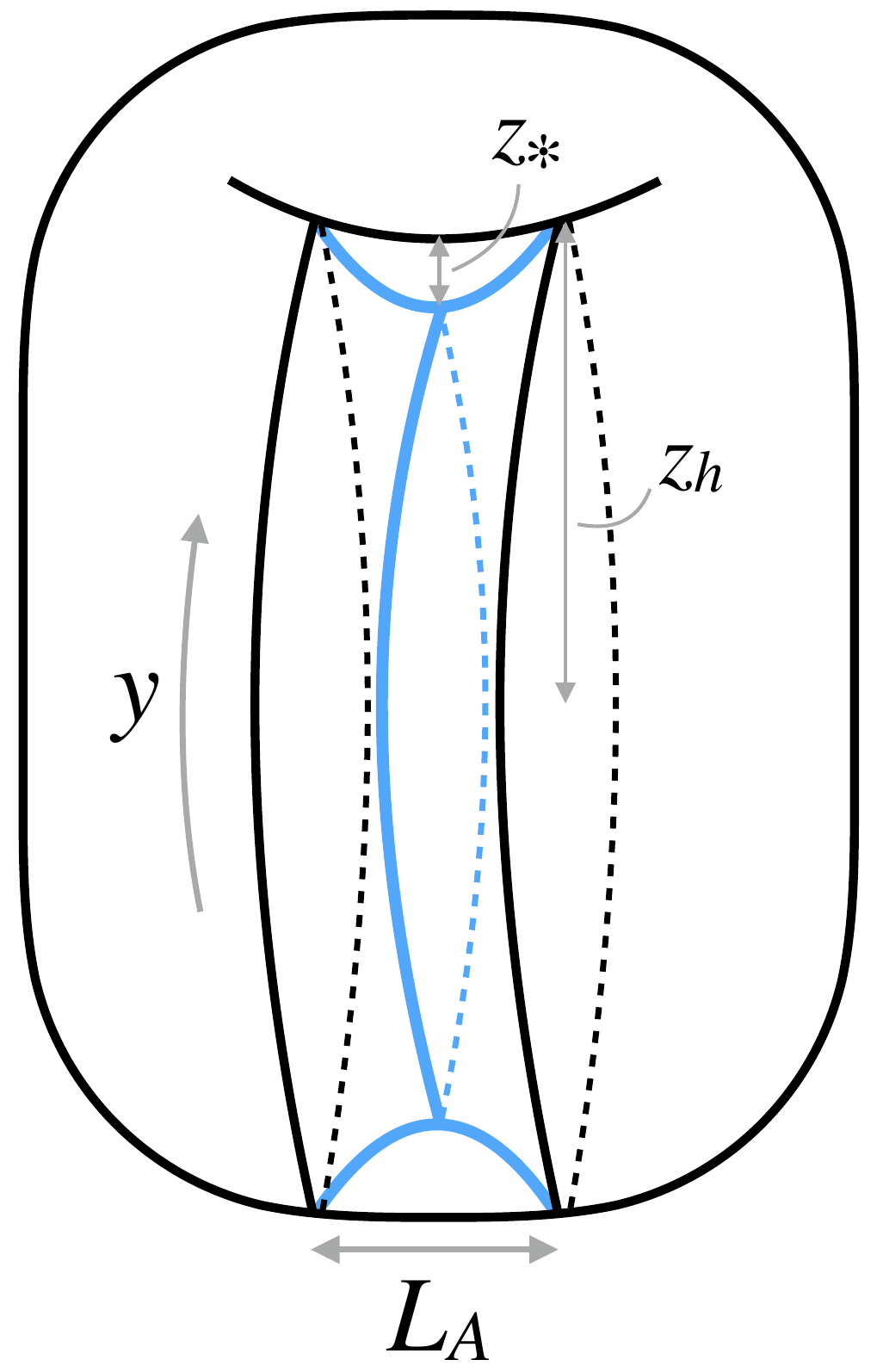}  
  \caption{Torus in 2 spatial dimensions for $L_y\geq L_x$, including a schematic representation of the minimal surface (blue) in the bulk.
    The holographic direction $z$ fills the inside of the torus, which represents an equal-time slice of the AdS-soliton.
    $z_*$ is the turning point of the minimal surface.
     The soliton spacetime pinches off at a ``horizon'' depth $z_h= 3L_x/(4\pi)$.
    The minimal surface remains connected when $L_y\geq L_x$.}   
  \labell{fat-soliton}
\end{figure} 

\subsection{Three dimensions}   
\label{sec:t3}   
In this section we study the holographic EE of a cylinder $A$ of size $L_A\times L_y$ wrapping one of the cycles of a 
rectangular 2-torus with aspect ratio $b=L_x/L_y$ --- see Fig. \ref{fig:torus3d}. In the first two subsections we review and extend   
the results of \cite{Chen14} on the properties 
of $\chi(\theta)$. In subsection \ref{cyl3} we study the limit in which the spatial manifold becomes an infinite cylinder. 
In that limit, when in addition $L_A/L_y\rightarrow \infty$, $\chi$ reduces to a constant, \emph{independent of any length scale} --- see \req{ees2}.  

As explained in section \ref{adsoli}, the relevant bulk solutions are the AdS$_4$ solitons \cite{Soli2,Soli1}:
\begin{align}\label{solito}
ds^2&=\frac{1}{z^2}\left[\frac{dz^2}{f}+g_{xx}\,dx^2+g_{yy}\,dy^2-dt^2 \right] ,
\end{align}
where we work in units where the AdS radius is set to unity, $L_{\rm AdS}=1$.
The spatial components of the metric read
\begin{align}\label{bl11}
  g_{xx}&=f(z)\, , \,\, g_{yy}=1\, , \quad \text{if} \quad b\leq1 \, ,\\ \label{bg1}
  g_{xx}&=1\, , \,\, g_{yy}=f(z)\, , \quad \text{if} \quad b\geq1 \, ,
\end{align}
with $f(z)=1-(z/z_h)^3$. Depending on whether $b$ is larger or smaller than $1$, the ground state of the system is described by the first solution or the second, respectively. This can be seen by comparing the energy densities of both solutions 
\begin{align}
\Delta \mathcal{E}=-\frac{4\pi^2 L_x L_y}{9G} \left[\frac{1}{L_x^3}-\frac{1}{L_y^3} \right]\, .
\end{align}
Whenever $1/L_x^3> 1/L_y^3$ \ie whenever $b\leq 1$, the first solution dominates the partition function, and vice versa. 

As explained in the previous subsection, the soliton geometries \req{solito} look like ``cigars'' in the $(x,z)$ and $(y,z)$ planes, 
respectively, with the tip being at $z\!=\! z_h$. 
Regularity at that locus imposes $z_h=3L_x /(4\pi)$ and $z_h=3L_y/(4\pi)$, respectively. 
This can be seen most easily by treating the contracting direction ($x$ or $y$) as Euclidean
time, and recalling that the time coordinate in a Wick-rotated AdS$_4$-Schwarzschild background is periodic with period
given by the inverse temperature,
$\beta=4\pi z_h/3$, where $z_h$ is the location of the black brane's horizon. Here the role $\beta$ is played by 
$L_x$ or $L_y$, depending on which one is the smallest.  
\subsubsection{$L_y\geq L_x$}\label{bl1}

Let us start considering the case in which the torus aspect ratio satisfies $b\leq 1$. As explained above, we need to consider the soliton geometry corresponding to \req{bl11}.
The holographic EE of the cylinder $A$ can be now computed using the Ryu-Takayanagi prescription \req{RyuTaka},
where the minimal surface is sketched in \rfig{fat-soliton}.   
The result reads \cite{Chen14}
\begin{equation}\label{ees}
S=\frac{L_y}{2G\epsilon}-\chi \, ,
\end{equation}
where $\epsilon$ is a short distance (UV) cut-off ($z\rightarrow \epsilon$ as we approach the boundary).
We note that area law term does not depend on the aspect ratio of the torus, nor on $L_A$. 
The universal term is
\begin{equation} \label{chite}
  \chi(\theta)=\frac{2\pi }{3b \xi^{1/3}\, G }\left[1 + \int_0^1\frac{-d\zeta}{\zeta^2}\left( \frac{1}{\sqrt{P(\xi,\zeta)}}-1\right) \right] \, ,
\end{equation}
where $P(\xi,\zeta)=1-\xi \zeta^3-(1-\xi)\zeta^4$, $\zeta=z/z_*$ and $\xi=(z_*/z_h)^3$. 
Here, $z_*$ is the value of $z$ corresponding to the turning point of the holographic surface, see Fig.~\ref{RT_cyl}. In the above expression we have written $\chi$ as a function of the ratio $\xi$, which is related to the \emph{angle} 
\begin{align}
  \theta = 2\pi \frac{L_A}{L_x}
\end{align}
through 
\begin{equation}\label{xite}
\frac{2\pi L_A}{L_x}=\theta(\xi)=3\xi^{1/3}\int_0^1\frac{d\zeta \,\zeta^2\sqrt{1-\xi}}{\sqrt{P(\xi,\zeta)}(1-\xi \zeta^3)}\, .
\end{equation}
Hence, in order to write $\chi$ as a function of $\theta$, one needs to invert \req{xite} and then substitute back $\xi(\theta)$ in \req{chite}.
Note that the dependence on the aspect-ratio factorizes out from $\chi(\theta)$, which seems to be a non-generic feature. 

In general, it is not possible to write down the function $\chi(\theta)$ explicitly. However, it is possible to find an analytic expansion around the so-called ``thin slice'' limit $\theta \rightarrow 0$: 
\begin{equation} \label{parox}
  \chi(\theta)=\left[\frac{2\pi\kappa}{b}\right]\frac{1}{\theta}+\left[\frac{   \Gamma(\frac{1}{4})^{12}}{1306368\pi^7 \,b \,G}\right]\theta^5 + \cdots\, ,
\end{equation} 
where  
\begin{equation}\label{parox2}
\kappa=\frac{2\pi^3}{\Gamma(\frac{1}{4})^4 G}
\end{equation}
is the quantity that controls the universal coefficient to the EE of a strip region in flat space, as discussed in general in Section~\ref{sec:basics}, see
Eq.~\req{thin-slice}.   
Indeed, in the $\theta\rightarrow 0$ limit, the torus EE reduces to that of a thin strip of width $L_A$ and length $L_y\gg L_A$. 
In our example, this can be verified holographically \cite{Chen14} since the strip EE is given by \cite{RyuTaka1} $S_{\rm strip}=L_y/(2G\epsilon)-\kappa L_y/L_A$,
with $\kappa$ exactly as in \req{parox2}.
Going beyond the leading term, $\chi(\theta)$ in our holographic CFT can be expanded as $\chi(\theta)=1/(b G)\sum_{k=0}a_k\, \theta^{3k-1}$,
where the $a_k$ are pure numbers. Interestingly, $a_1=0$ whereas $a_0$ and $a_2$ are given in \req{parox}.  
The 2-term expansion \req{parox} fits the exact function (which we can only access numerically) very accurately even for intermediate values of $\theta$, 
as shown in Fig.~\ref{fig1}. 
\begin{figure}
  \centering
  \includegraphics[scale=0.221]{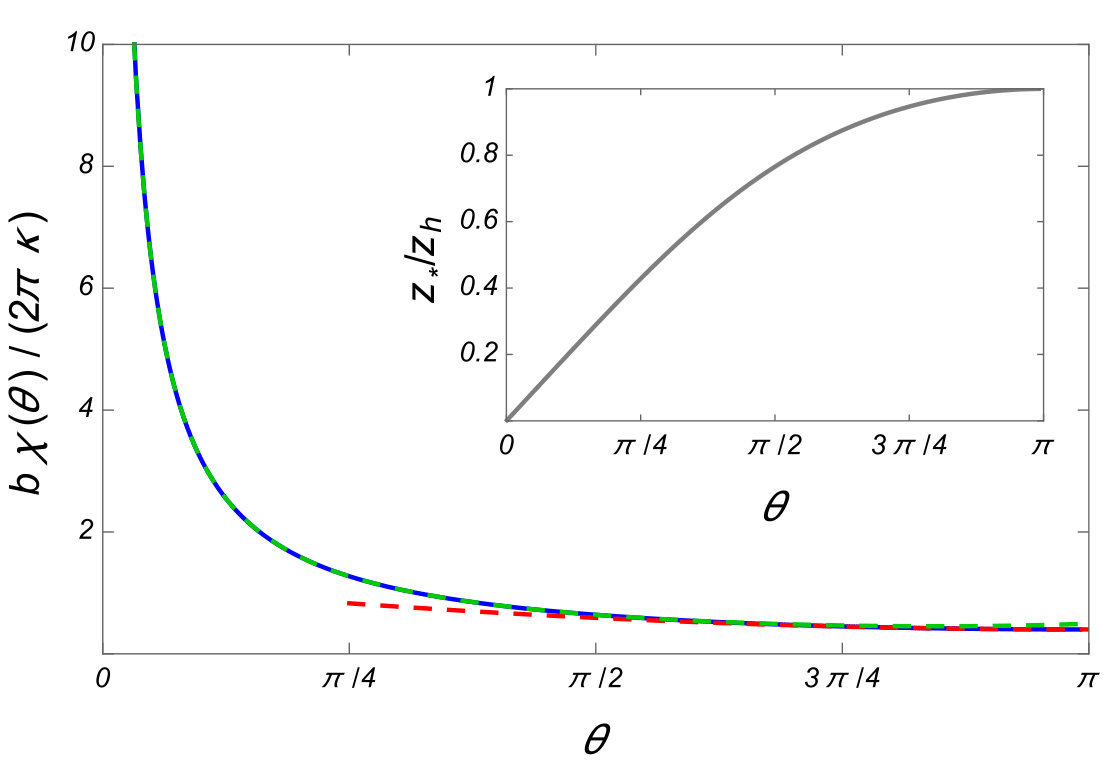}
  \caption{\textbf{Main}: $\chi(\theta)b /(2\pi \kappa)$ as a function of $\theta$ for $b\leq 1$. The green dashed line corresponds to the first two terms of the small $\theta$ approximation in \req{parox}. The red dashed line corresponds to the first two terms of the 
    expansion around $\theta\!=\!\pi$, Eq.~\req{smooth-lim}. 
Observe that the factorization of $b$ from $\chi(\theta)$ implies that the normalized 
curve $\chi(\theta)b /(2\pi \kappa)$ is the same for all values of $b\leq 1$. 
\textbf{Inset}: Ratio $z_*/z_h$ as a function of $\theta$.}  
  \labell{fig1}
\end{figure} 

In the opposite limit where region $A$ is of maximal length, $\theta\to \pi$, the minimal surface fills the entire soliton, $z_*\to z_h$, and we find
\begin{equation}\label{tetis}
  \chi(\pi)=\frac{2\pi^{3/2}\Gamma(\frac{2}{3})}{3\Gamma(\frac{1}{6})} \, \frac{1}{b\, G} \,,
\end{equation} 
which yields approximately $0.90307/(bG)$.    
We recall from Section~\ref{sec:basics} that $\chi(\theta)$ is an even analytic function around $\pi$.
The expansion about this limit is thus $\chi=\chi(\pi)+c_1\cdot (\theta-\pi)^2+\dotsb$, as given in Eq.~\req{smooth-lim}. We have analytically obtained
the coefficient of the quartic term by expanding \req{chite}:
\begin{align}
c_1=\frac{\pi}{18b\, G}\, .
\end{align}
We stress that $c_1$ is always non-negative, as required by strong subadditivity of the EE \cite{will-torus}.   

\subsubsection{$L_x\geq L_y$}\label{bg11}  

As observed in \cite{Chen14}, the situation is more subtle when $b\geq 1$. This is because of the existence of a 
disconnected holographic minimal surface that becomes minimal beyond a certain value of the ratio $L_A/L_y$.
This connected-to-disconnected transition is illustrated in \rfig{RT_cyl}. The transition will have important consequences
on the EE \cite{Chen14}, as we now review. 
\begin{figure}
  \centering
  \includegraphics[scale=0.222]{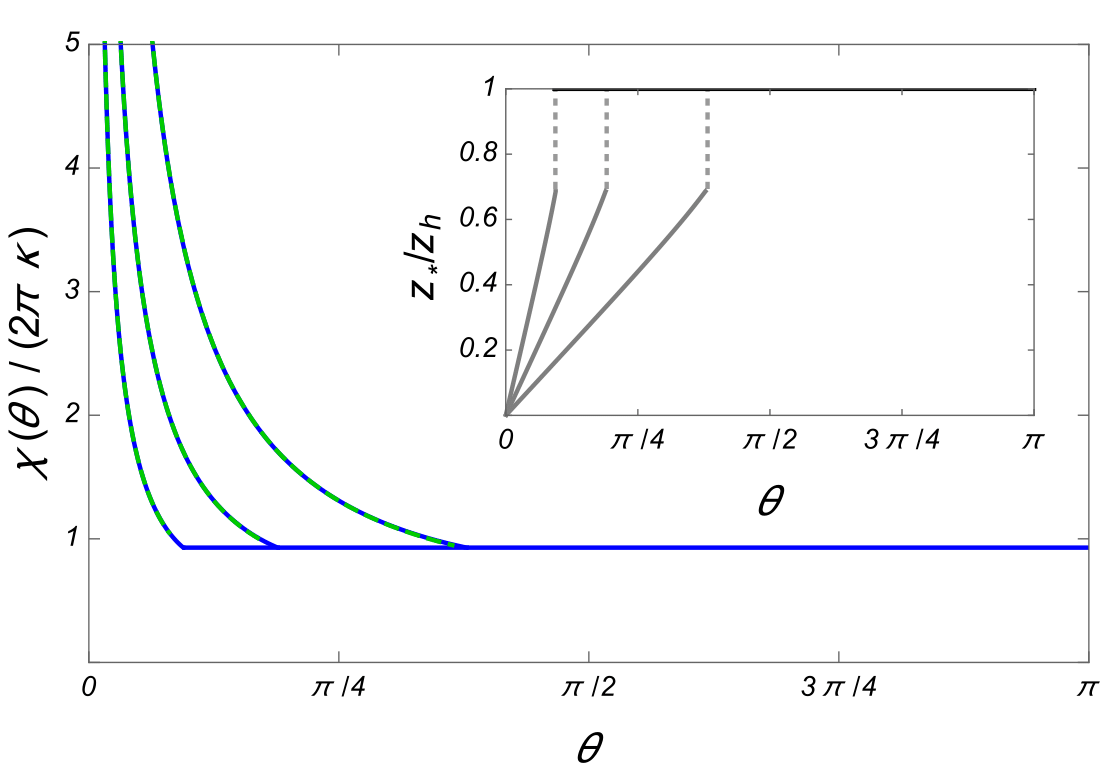}
  \caption{\textbf{Main}:  
    $\chi(\theta) /(2\pi \kappa)$ as a function of $\theta$ for $b=1^+,2,4$ (from right to left) and the saturation value: $\chi(\theta) /(2\pi \kappa)=\Gamma[1/4]^4/(6\pi^3)$. The green dashed lines correspond to the first two terms of the small $\theta$ approximation in \req{paroox}. \textbf{Inset}: Ratio $z_*/z_h$ as a function of $\theta$ (also from right to left). 
    At $\theta=2\pi p /b\simeq 1.1867/b$, the turning point $z_*$ jumps to $z_h$.}  
  \labell{fig2}
\end{figure} 
The final result for the holographic EE is again given by \req{ees},
where now \cite{Chen14}
\begin{align}\label{casi}
  \chi(\theta)&=\tilde{\chi}(\theta)\, , \qquad 0<\frac{\theta}{2\pi} \leq \frac{p}{b}\, , \\ 
  \chi(\theta)&=\frac{2\pi}{3G}\, , \qquad \frac{p}{b}<\frac{\theta }{2\pi} \leq \frac{1}{2}\, , \notag 
\end{align}
where 
\begin{align}\label{ttt}
  \tilde{\chi}(\theta)=\frac{2\pi }{3\, G \xi^{1/3}}\left[\int_0^1\frac{-d\zeta}{\zeta^2}\left[ \frac{\sqrt{1-\xi \zeta^3}}{\sqrt{P(\xi,\zeta)}}-1\right]+1\right]\, ,
\end{align}
and the ratio $\xi$ is related to $\theta$ through
\begin{equation}\label{lali}
  \frac{2\pi L_A}{L_x}=\theta(\xi)=\frac{3\xi^{1/3}}{b}\int_0^1\frac{d\zeta \,\zeta^2\,\sqrt{1-\xi}}{\sqrt{P(\xi,\zeta)(1-\xi \zeta^3)}}\,.
\end{equation}  
$\chi$ for $\pi<\theta <2\pi$ is obtained by using the reflection property $\chi(2\pi-\theta)=\chi(\theta)$. 
We have defined the numerical constant  
\begin{align}
  p\simeq 0.1889\, .
\end{align}
Hence, when $L_A = p L_y$ there is an abrupt transition and $\chi(\theta)$ stops changing 
as $L_A$, \ie $\theta$, increases further. $\chi$ is shown in Fig.~\ref{fig2} for 3 aspect ratios, $b=1^+, 2, 4$. 
The reason for the observed saturation is that the minimal surface becomes  
the sum of two disconnected surfaces, each one of which would correspond to the entangling surface   
for a semi-infinite cylindrical entangling region --- see next subsection for a discussion of this specific limit.
The torus EE $\chi$ thus obtained in holography is singular: not only is there a cusp as a function of $\theta$ at the transition point,
but also $\chi(\theta;b)$ is discontinuous at $b=1$,  taking different values at $b=1^-$ (Fig.~\ref{fig1}) and $1^+$ (Fig.~\ref{fig2}).
This singular behavior is not generic, and for instance is absent in the free scalar \cite{will-torus,PSI,WWS16} and Dirac fermion \cite{Chen14} CFTs,
at the large-$n$ O$(n)$ Wilson-Fisher fixed point \cite{WWS16}, 
and in the so-called Extensive Mutual Information model \cite{will-torus}. It would be interesting
to see to what extent it becomes smoothed out by including $1/N$ corrections.

Observe that, as opposed to the $b\leq 1$ case, now the dependence on $b$ appears through $\theta(\xi)$ and not through the dependence of $\tilde{\chi}$ on $\xi$. This implies, in particular, that $b$ does not factorize from $\tilde{\chi}(\theta)$.   

We now examine the thin slice limit, $\theta\to 0$, where we find
\begin{equation}\label{paroox}
  \chi(\theta)=\left[\frac{2\pi \kappa}{b}\right]\frac{1}{\theta}+\left[\frac{   \Gamma(\frac{1}{4})^4 b^2}{432\pi\,G}\right]\theta^2 + \cdots
\end{equation}  
Observe that while the leading term coincides with the $b\leq 1$ case in \req{parox}, as expected, the 
subleading contribution is of order $\theta^2$ instead of $\theta^5$. The full series can be now written as $\tilde{\chi}(\theta)=\sum_{k=0}a_k\, (b\theta)^{3k-1}$, where the $a_k$ are pure numbers. As we can see from Fig.~\ref{fig2}, the approximation given by \req{paroox} fits the exact curve with great accuracy for all values of $\theta < 2\pi p/b$ in each case. 

\begin{figure}[h] 
  \centering
  \includegraphics[scale=0.4]{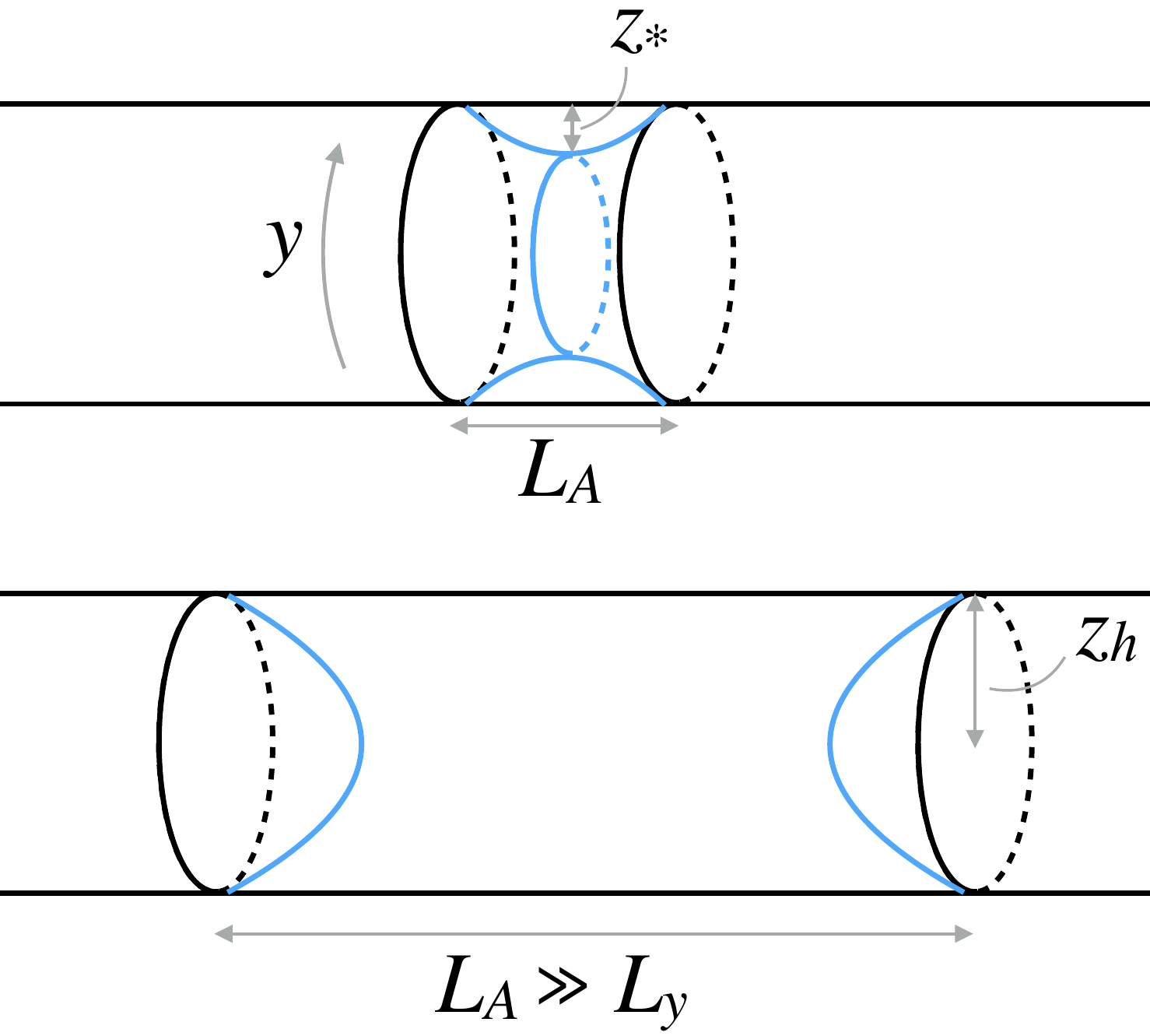} 
  \caption{Schematic representation of the minimal surface (blue) in the bulk, in the limit where 
    the spatial manifold is a thin torus, $L_x\gg L_y$. We only show the portion near region $A$, which is a cylinder of length $L_A\ll L_x$.
    The holographic direction $z$ fills the inside of the cylinder. 
    $z_*$ is the turning point of the minimal surface.
     The soliton spacetime pinches off at a depth $z_h=3L_y/(4\pi)$.
    We note that the minimal surface undergoes a transition from being connected when $L_A< p L_y$ (top), to
being disconnected when $L_A>p L_y$ (bottom).}   
  \labell{RT_cyl}
\end{figure}

\subsubsection{Infinite cylinder \& thin torus limits} \label{cyl3}

When $L_x\to \infty$, the torus becomes an infinitely long cylinder, and the EE only depends on $L_A/L_y$. 
This limit was previously studied for the free scalar CFT \cite{Arias2015}, and for the Extensive Mutual Information model \cite{will-torus}.
The expansion for the full EE $S$, \req{ees}, remains valid and needs to use \req{lali} to express $\xi$ 
as a function of $L_A/L_y$ instead of $\theta$, \ie
\begin{equation}\label{lalii}
\frac{2\pi L_A}{L_y}=3\xi^{1/3}\int_0^1\frac{d\zeta \,\zeta^2\,\sqrt{1-\xi}}{\sqrt{P(\xi,\zeta)(1-\xi \zeta^3)}}\, .
\end{equation}
Then, $\chi(L_A/L_y)$ is given by 
\begin{align}\label{casii}
\chi(L_A/L_y)&=\tilde{\chi}(L_A/L_y)\, , \qquad 0<\frac{L_A}{L_y}< p\, , \\ \notag
\chi(L_A/L_y)&=\frac{2\pi}{3G}\, , \qquad p<\frac{L_A}{L_y}< \infty \, ,
\end{align}
where $\tilde{\chi}$ is again given by \req{ttt}, but now with $\xi(L_A/L_y)$ computed using \req{lalii}.  
Hence, $\chi(L_A/L_y)/(2\pi \kappa)$ exactly corresponds to the $b=1$ curve in Fig. \ref{fig2}, with the difference that the saturation regime would extend all the way to $L_A/L_y\rightarrow \infty$. 
In this limit where region $A$ becomes very long, $\chi\to 2\gamma$, where $\gamma$ is a universal constant \emph{independent of any 
length scale.}
The factor of 2 comes because region $A$ is a finite cylinder with two boundaries.
In our holographic CFT, $\gamma=\pi/(3G)$.
More simply, one can consider the new situation where space is an infinite cylinder of circumference $L_y$ 
and region $A$ is a semi-infinite cylinder,
in which case the EE becomes
\begin{equation}\label{ees2}
S=\frac{L_y}{4G\epsilon}-\gamma\, . 
\end{equation}
With respect to \req{casii}, this contains an additional factor $1/2$ because here $A$ has a single boundary. 

Observe that the structure of \req{ees2} is very similar to the EE of a radius-$R$ disk in a three-dimensional CFT in flat space, which reads $S_{\rm disk}=\mathcal{B} R/\epsilon - F$, where $F$ is independent of both the regulator and the disk's radius --- when computed with sufficient care \cite{Casini:2015woa}.

\subsection{Four dimensions}   \label{sec:torus4}

We now turn to the case of a four-dimensional boundary theory. We will study the EE of a cylinder $A$ with dimensions 
$L_A\times L_1 \times L_2$ wrapping two of the cycles of the spatial manifold,
a 3-torus $\mathbb T^3=L_x\times L_1 \times L_2$,  
with aspect ratios $b_i=L_x/L_i$, $i=1,2$ --- see Fig.~\ref{fig:torus4d}. The first two subsections will be devoted to the general 
properties of $\chi(\theta)$. In subsection \ref{cylicio}, we will consider the case in which $L_x$ becomes infinite with special emphasis on the 
$L_A/L_{1,2}\to \infty$ limit, for which $\chi(\theta)$ becomes a function of the single ratio, $L_2/L_1$, 
\ie $\chi\to 2\gamma(L_2/L_1)$. 

\begin{figure}
        \centering
                \includegraphics[scale=0.4]{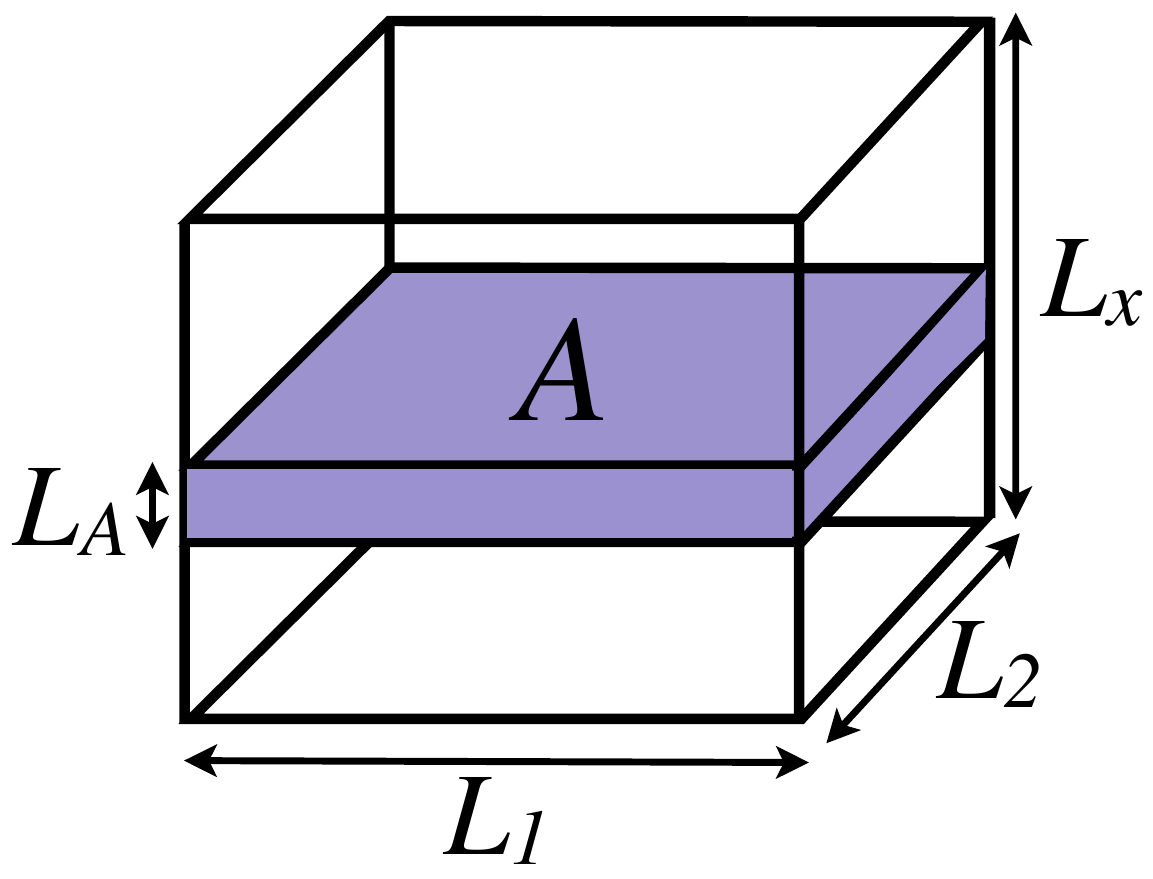}
                \caption{Flat torus in 3 spatial dimensions; opposite faces are identified. The entangling region $A$
is the cylinder along the $x$-direction. Each of the 2 boundaries of $A$ is a 2-torus, $\mathbb T^2$.}  
\labell{fig:torus4d}
\end{figure}

As explained in the introduction, we need to consider AdS$_5$ solitons, whose metrics are given by
\begin{align}  \label{soliti}
  ds^2 = \frac{1}{z^2}\left[\frac{dz^2}{f}+g_{xx}dx^2+g_{11}(d{y^1})^2+g_{22}(d{y^2})^2-dt^2 \right]\! , 
\end{align}
where the components of the metric read
\begin{align}\label{bg0}
g_{xx}&=f\, , \,\, g_{11}=1\, , \,\, g_{22}=1\, , \quad \text{if} \quad L_1,L_2\geq L_x \, ,\\ \label{bg1}
g_{xx}&=1\, , \,\, g_{11}=f\, , \,\, g_{22}=1\, , \quad \text{if} \quad L_x,L_2\geq L_1 \, , \\ \label{bg2}
g_{xx}&=1\, , \,\, g_{11}=1\, , \,\, g_{22}=f\, , \quad \text{if} \quad L_x,L_1\geq L_2 \, .
\end{align}
The blackening factor is in turn given by $f(z)=1-(z/z_h)^4$. In the cases \req{bg0}, \req{bg1} and \req{bg2}, regularity at the tip of the cigar imposes respectively $z_h=L_x/\pi$, $z_h=L_1/\pi$, and $z_h=L_2/\pi$, respectively. 
Just like in the $d=3$ case, the conditions \req{bg0}, \req{bg1} and \req{bg2} correspond to the solutions with the 
smallest energy density in each case. Explicitly, one finds 
\begin{equation}
\mathcal{E}=-\frac{ L_1 L_2 \pi^3}{16 G L_x^3 }\, , \quad \mathcal{E}=-\frac{ L_x L_2 \pi^3}{16 G L_1^3 }\, , \quad \mathcal{E}=-\frac{ L_1 L_x \pi^3}{16 G L_2^3 }\, ,
\end{equation}
respectively for \req{bg0}, \req{bg1} and \req{bg2}. 

Observe that in principle there is no reason to treat the $x$ direction differently from $y^1$ and $y^2$. 
The reason why we do so is that $A$ covers the $y^1$ and $y^2$ directions entirely, but not $x$. In particular, this will imply that the cases \req{bg1} and \req{bg2} will be equivalent --- the only modification in the results 
being the switch $L_1 \leftrightarrow L_2$ --- but of course different from the one corresponding to \req{bg0}.

\subsubsection{$L_{1},\, L_{2}\geq L_x$}
Let us start with the case $b_1,b_2\leq 1$, for which we need to pick \req{bg0}. Using the RT prescription \req{RyuTaka}, one finds the following expression for the holographic EE of the cylindrical region $A$,
\begin{equation}\label{tete}
S= \frac{ L_{1} L_{{2}}}{4G }\frac{1}{\epsilon^{2}}-\chi(\theta) \, . 
\end{equation}
The first term is the boundary law contribution, proportional to the area of each $\mathbb T^2$ boundary of $A$, $L_1L_2$.
Being local in nature, it is independent of the aspect ratios $b_{1,2}$ and of $L_A$.
The universal subleading term is 
\begin{equation} \label{chi4d}
  \chi(\theta)=\frac{\pi^{2}  }{4b_1 b_{2} \xi^{1/2}G}\left[\int_0^1\frac{-2d\zeta}{\zeta^{3}}\left[ \frac{1}{\sqrt{P(\xi,\zeta)}}-1\right]+1\right], 
\end{equation} 
with $\xi=(z_*/z_h)^4$, $\zeta=z/z_*$ and where now $P(\xi,\zeta)=1-\xi \zeta^4-(1-\xi)\zeta^{6}$. Again, $z_*$ is the value of $z$ corresponding to the turning point of the holographic surface. In this case, the dependence of $\chi$ 
on the angle $\theta$ can be obtained from  
\begin{equation}
  \frac{2\pi L_A}{L_x}=\theta(\xi)=4 \xi^{1/4}\int_0^1\frac{d\zeta \,\zeta^{3}\sqrt{1-\xi}}{\sqrt{P(\xi,\zeta)}(1-\xi \zeta^4)}\, .
\end{equation}  
From \req{chi4d}, we note that $b_1$ and $b_2$ factorize from $\chi(\theta)$, in a similar fashion to the AdS$_4$ case.

At this point it is worth recalling that in the standard discussion of EE in 4d CFTs, 
where the spatial manifold is infinite flat space $\mathbb R^3$,
the subleading contribution associated with a generic smooth surface has a logarithmic dependence on the cut-off: 
$S_{\rm univ}= I(a,c) \log(l/\epsilon)$. $I(a,c)$ is universal ($\epsilon$-independent) 
and consists of an integral over the entangling surface $\partial A$ of certain geometric quantities controlled 
by the central charges $a$ and $c$ \cite{Solodukhin:2008dh}. 
Note however that if the intrinsic and extrinsic curvatures of $\partial A$ vanish, $I(a,c)$ is zero, 
and the subleading contribution to the EE will now be a universal constant instead. This is the case when $A$ is an infinite 
flat slab, for instance.
In our case, $A$ can be understood as a slab in flat space with the peculiarity that the spatial dimensions have been compactified. 
As a consequence, the universal contribution in \req{tete} is constant, and does not grow logarithmically with the cutoff. 

Again, it does not seem possible to write a general explicit expression for $\chi(\theta)$ in terms of basic functions. 
We can nevertheless analytically obtain the leading term in the thin slice $\theta\rightarrow 0$ limit. In this regime, $\chi(\theta)$ admits an expansion of the form
\begin{equation} \label{paroox4}
  \chi(\theta)=\frac{1}{b_1 b_2 G}\left[\frac{4\pi^{7/2}\Gamma(\frac{2}{3})^3}{\Gamma(\frac{1}{6})^3 }\frac{1}{\theta^2}+ \sum_{k=1}a_k\, \theta^{4k-2}\right] ,
\end{equation}
where the $a_k$ are pure numbers, the first of which reads $a_1\simeq -0.116305$. 
As shown in Fig.~\ref{fig3}, the first two terms in \req{paroox4} fit the exact curve rather accurately for reasonably large values of $\theta$. 

\begin{figure}
        \centering
                \includegraphics[scale=0.22]{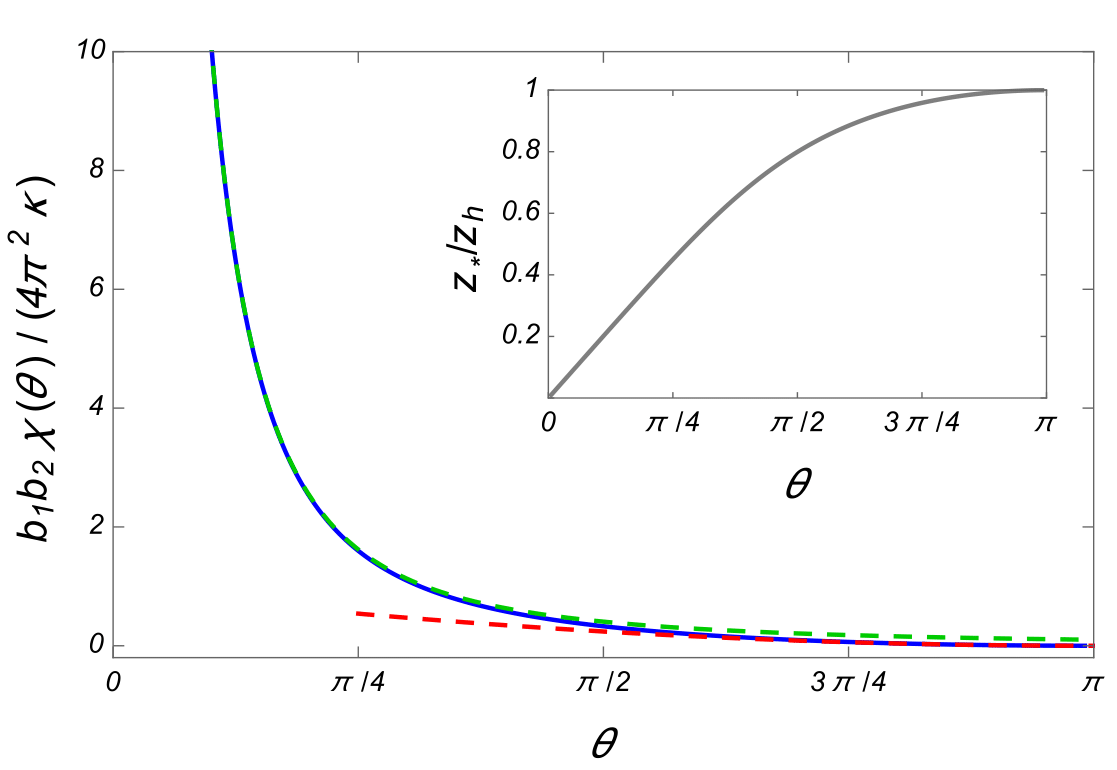}
                \caption{\textbf{Main}: 
                $ \chi(\theta) b_1 b_2/(2\pi \kappa)$ as a function of $\theta$. The green dashed line corresponds to the first two terms of the small $\theta$ approximation in \req{paroox4}. The red dashed line is the leading term in the expansion around $\theta=\pi$, \req{enen}. 
Observe that in this case $\chi(\pi)\!=\!0$. \textbf{Inset}: Ratio $z_*/z_h$ as a function of $\theta$. } 
\labell{fig3}
\end{figure}
\begin{figure}
  \centering
  \includegraphics[scale=0.22]{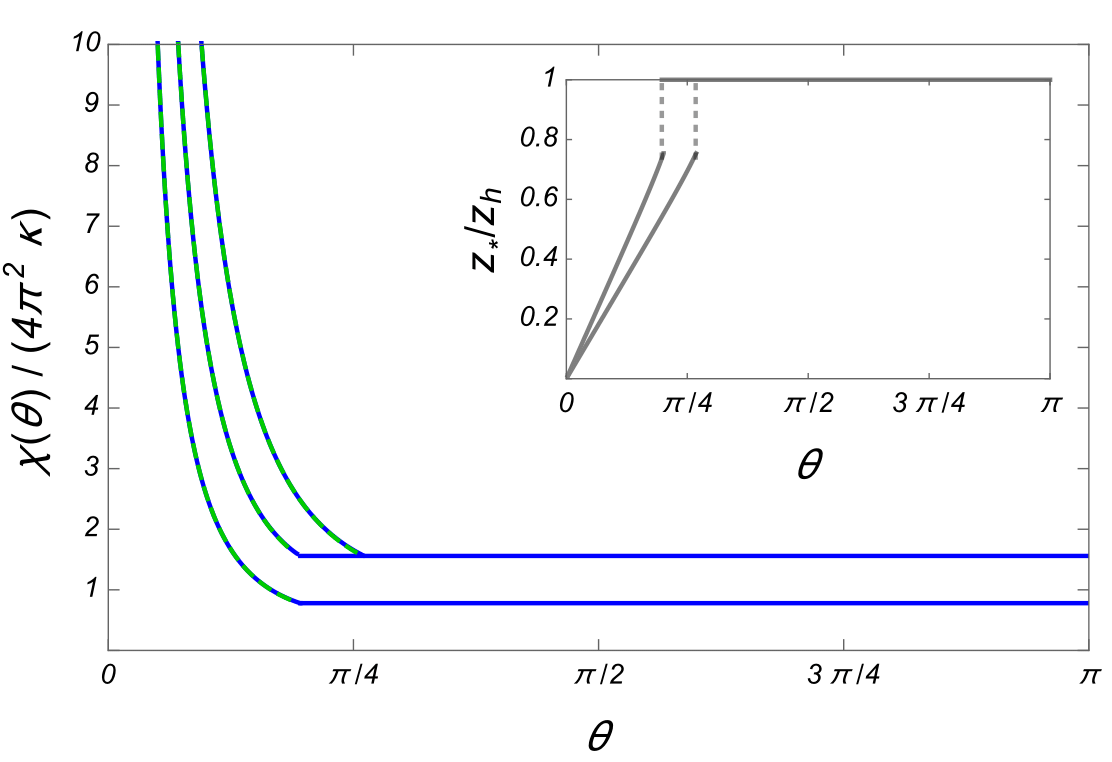}
  \caption{\textbf{Main}: 
    $\chi(\theta) /(4\pi^2 \kappa)$ as a function of $\theta$ for $b_1=3/2$, $b_2=3/4$; $b_1=2$, $b_2=1$; 
$b_1=b_2=2$; (from right to left) and the saturation values $\chi(\theta)/(4\pi^2\kappa)=\Gamma(1/6)^3b_1/(16 \pi^{3/2}\Gamma(2/3)^3 b_2)$.  
The green dashed lines are the small-$\theta$ approximations in \req{paroox4} including the 
first 2 terms. 
\textbf{Inset}: $z_*/z_h$ as a function of $\theta$. The right curve corresponds to $b_1=3/2$, $b_2=3/4$, while 
the left one to  $b_1=2$, $b_2=1$ and $b_1=b_2=2$. At $\theta=2\pi p /b_1\simeq 1.3/b_1$, respectively, the functions jump to the value $z_*/z_h=1$.} 
  \labell{fig4}
\end{figure}

Note that the leading term in \req{paroox4} was to be expected from the fact that, in this limit, $A$ becomes 
a thin slab with thickness $L_A$, whose EE is given by $S_{\rm strip}=L_1 L_2/(4G\epsilon)-\kappa L_1 L_2/L_A^2$ \cite{RyuTaka1}, where
\begin{equation}
  \kappa=\frac{\pi^{3/2}\Gamma(\frac{2}{3})^3}{\Gamma(\frac{1}{6})^3 G}\, . 
\end{equation}
Quite remarkably, $\chi(\theta)$ exactly vanishes when $L_A$ is half of $L_x$, \ie when $\theta=\pi$,
\begin{equation}\label{c44}
\chi(\pi)=0\, .
\end{equation}
Observe that this behavior differs from the three-dimensional case, \req{tetis}, for which a positive value is found. 
Around that limit, $\chi(\theta)$ behaves as 
\begin{equation}\label{enen}
\chi(\theta)=\sum_{\ell=1} c_{\ell}\cdot (\pi-\theta)^{2\ell}\, ,
\end{equation}
where the leading coefficient reads in this case
\begin{equation}
c_1=\frac{\pi^2}{32b_1b_2 G}\,.
\end{equation}
Just as in $3d$, this coefficient is always non-negative, as required by strong subadditivity of the EE \cite{will-torus}.  

\subsubsection{$L_x,\,L_{2}\geq L_{1}$} \label{tortak}   
Let us now consider the case given by \req{bg1}, corresponding to $b_1\geq b_2, 1$, \ie $L_1$ is smaller than $L_2$ and $L_x$.
Some of the results of this section in the limit $L_x\to \infty$ were previously presented in \cite{Takabu}, where the authors 
were interested --- among other things --- in the interpretation of the AdS$_5$ soliton as a holographic dual of 
certain four-dimensional non-supersymmetric gapped Yang-Mills theories \cite{Soli2}. 
Indeed, the AdS-soliton background that we study corresponds to anti-periodic boundary conditions for the fermions,
leading to a breaking of supersymmetry.  

The result for the EE of a cylindrical region $A$ is again given by \req{tete}, where now   
\begin{align}
\chi(\theta)&=\tilde{\chi}(\theta)\, , \qquad 0<\frac{\theta}{2\pi}\leq \frac{ p}{b_1}\, , \\ \label{tiwy}
\chi(\theta)&=\frac{b_1\pi^2}{4b_2 G}\, , \qquad \frac{p}{b_1}<\frac{\theta }{2\pi}\leq \frac{1}{2}\, .
\end{align}
Again, $\chi$ for $\pi<\theta <2\pi$ is obtained by using the reflection property $\chi(2\pi-\theta)=\chi(\theta)$.
Here 
\begin{align}
  p \simeq 0.1958 \, ,
\end{align}
--- a value which is quite close to the $d=3$ one --- and
\begin{align}\label{coshs}
\tilde{\chi}(\theta)&=\frac{\pi^{2} b_1 }{ 4\xi^{1/2} b_{2} G}\left[\int_0^1\frac{-2d\zeta}{\zeta^{3}}\left[ \frac{\sqrt{1-\xi \zeta^4}}{\sqrt{P(\xi,\zeta)}}-1\right]+1\right]\, ,\\
\label{tititi}
\frac{2\pi L_A}{L_x}&=\theta(\xi)=\frac{4\xi^{1/4}}{b_1} \int_0^1\frac{d\zeta \,\zeta^{3} \sqrt{1-\xi}}{\sqrt{P(\xi,\zeta)(1-\xi \zeta^4)}}\, .
\end{align}
Observe that when $L_A =\,p\, L_1$, a phase transition similar to the one found in $d=3$ occurs in $\chi(\theta)$. Indeed, when $L_A$ is larger than such value, the minimal surface no longer connects the two boundaries of $A$. The reason is the same, namely after that 
point one is left with two disconnected surfaces each one of which would correspond to the 
entangling surface for a semi-infinite cylindrical entangling region. 
We expect that the non-smoothness of $\chi(\theta)$ is partly an artifact of the large-$N$ limit of the present holographic theory.  
For instance, in the free boson CFT in $d=4$, $\chi$ varies smoothly as a function of both $\theta$  
and $b$ \cite{will-torus}, which is also the case in the Extensive Mutual Information Model \cite{will-torus}.   

Note that even though the value of $\theta$ for which the saturation occurs depends only on $b_1$, the value that $\chi(\theta)$ takes at that point depends on the ratio $b_1/b_2$. Hence, for instance, the saturation value for $b_1=3/2$, $b_2=3/4$ coincides with the one corresponding to $b_1=2$, $b_2=1$, while the value of $\theta$ at which the latter saturates is the same as for $b_1=b_2=2$, namely $\theta=\pi p$ --- see Fig. \ref{fig4}.

Observe also that for small values of $\theta$, $\tilde{\chi}(\theta)$ satisfies 
\begin{align} \label{paroox41}
  \tilde{\chi}(\theta)=\frac{b_1}{b_2G}\left[\frac{4\pi^{7/2}\Gamma(\frac{2}{3})^3}{\Gamma(\frac{1}{6})^3 b_1^2}\frac{1}{\theta^2}+ \sum_{k=1}a_k\, (b_1\theta)^{4k-2}\right], 
\end{align} 
where the coefficient of the subleading term reads $a_1\simeq 0.23261$. Again, the $a_k$ 
are pure numbers. As expected, the leading term in \req{paroox41} coincides with the one in \req{paroox4}.
The agreement between the first two terms in \req{paroox41} and the exact curve is extremely good in the whole range, as shown in Fig. \ref{fig4}.

As we explained before, the case $L_x,L_1\geq L_2$ is completely equivalent to the one considered here (replace $1\leftrightarrow 2$ everywhere). 

\subsubsection{Infinite cylinder \& thin torus limits} \label{cylicio} 
We consider a special limit of the full torus function, in which the spatial manifold becomes an infinite cylinder $\mathbb R\times \mathbb T^2$,
where $L_x\to\infty$ while $L_{1,2}$ remain finite. 
We first assume $L_2\geq  L_1$.
Analogously to the three-dimensional case, \req{tete} remains valid, while we need to replace \req{tititi} by
\begin{equation}\label{tititi2}
  \frac{2\pi L_A}{L_1}=4\xi^{1/4}\int_0^1\frac{d\zeta \,\zeta^{3} \sqrt{1-\xi}}{\sqrt{P(\xi,\zeta)(1-\xi \zeta^4)}}\, ,
\end{equation} 
so we can write $\xi$ as a function of $L_A/L_1$. Then, $\chi(L_A/L_1)$ reads
\begin{align}
\chi(L_A/L_1)&=\tilde{\chi}(L_A/L_1)\, , \qquad 0<\frac{L_A}{L_1}< p\, , \\ 
\chi(L_A/L_1)&=\frac{\pi^2 L_2}{4L_1 G}\, , \qquad p<\frac{L_A }{L_1} < \infty\, ,  
\end{align}
where $\tilde{\chi}(L_A/L_1)$ is given by \req{coshs} with the exception that $\xi(L_A/L_1)$ should be now computed using \req{tititi2}. 
Note that the function $\chi(L_A/L_1)$ is identical to $\chi(\theta)$ 
if in the latter we make the replacements  $b_1\rightarrow1$ and $b_2\rightarrow L_1/L_2$, with the constant value region extending all the way to $L_A/L_1= \infty$. 
\begin{figure}[h]
        \centering
                \includegraphics[scale=0.21]{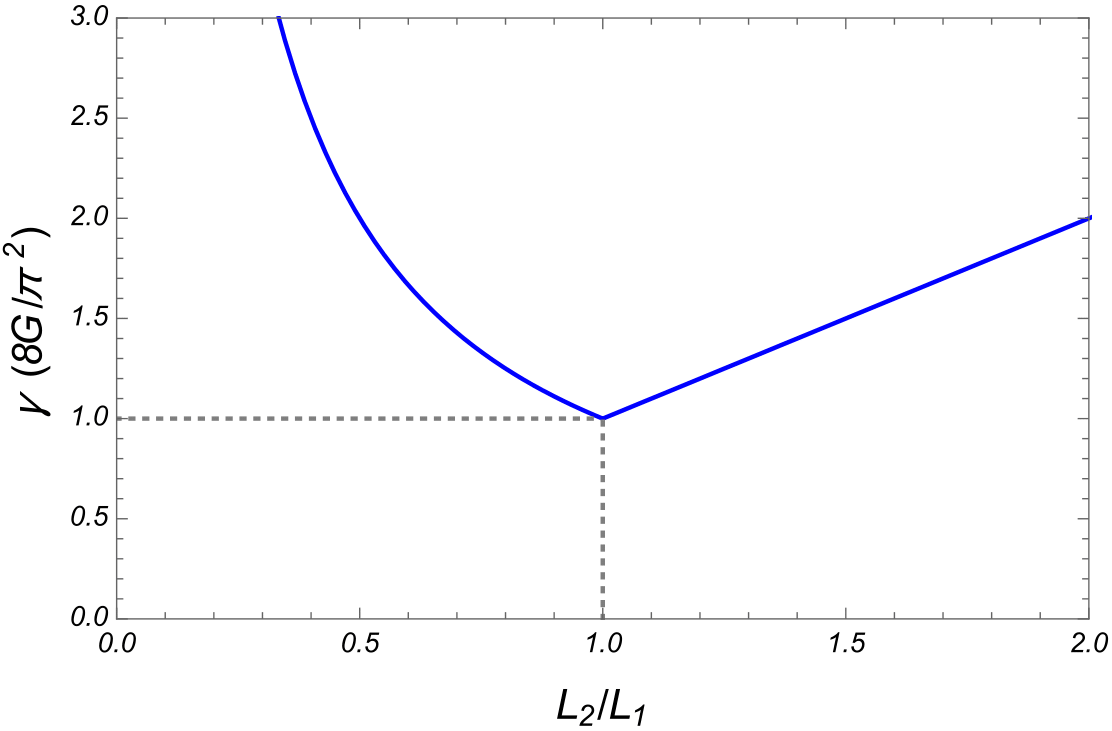}
                \caption{$\gamma \cdot(8G/\pi^2)$ as a function of $L_2/L_1$. There is an abrupt change at $L_1=L_2$.} 
\labell{fig5}
\end{figure}
In that limit, $\chi \rightarrow 2\gamma$, where 
\begin{equation}\label{gamba}
\gamma=\frac{\pi^2 }{8G}\, \frac{L_2}{L_1}\,, \quad L_2 \geq L_1\,,
\end{equation}
is a positive function of the dimensionless ratio $L_2/L_1$. 
One could obtain $\gamma$ more directly by considering region $A$ to be a semi-infinite cylinder 
spanning half of $\mathbb R\times \mathbb T^2$, in which case the EE becomes 
\begin{equation}\label{ees33}
S=\frac{L_1L_2}{8G\epsilon^2}-\gamma\,,  
\end{equation}
without the factor of 2 because $A$ has a single boundary. 
The above expression was obtained for the first time in \cite{Takabu}, where it was conjectured that the geometry \req{soliti} with \req{bg2} minimizes the value of $-\gamma$ within the class of locally asymptotically AdS$_5$ geometries. 

Observe that we are considering the case $L_2\geq L_1$. When $L_1\geq L_2$ instead, $\gamma$ naturally reads $\gamma=\pi^2 L_1/(8L_2 G)$. This means that the dependence of $\gamma$ on the quotient $L_2/L_1$ is linear, 
$\gamma \propto L_2/L_1$, for $L_2\geq L_1$, while it is, $\gamma \propto 1/(L_2/L_1)$, when $L_1\geq L_2$. 
We plot such behavior in Fig.~\ref{fig5}. The function is continuous everywhere, but non-differentiable at $L_2=L_1=L$.
For that value, the dependence on $L$ disappears from $\gamma$, resembling the three-dimensional result \req{ees2} 
\begin{equation}\label{ees3d3}
S=\frac{L^2}{8G\epsilon^2}-\frac{\pi^2}{8G}\, .
\end{equation}


\section{Renormalized entanglement entropies \& RG} \label{reer}

We now turn to the RG flow of the EE on the torus. As a theory flows away from its CFT fixed point, one is faced 
with the challenge of characterizing the EE in a universal way, \ie independent of the short-distance cutoff $\delta$. 
Indeed, the naive subtraction of the area law is no longer  
reliable \cite{Liu2012}, since the UV cutoff entering in the calculation of the EE can also flow.     
In \cite{Liu2012} (see also \cite{CH_F}), a useful notion of \emph{renormalized entanglement entropy} (REE) was introduced for general QFTs 
living in $d$-dimensional Minkowski space, $\mathbb R^{1,d-1}$. This REE thus not directly apply to our situation 
since our spatial manifold is a torus instead of the infinite plane.  
After reviewing the Minkowski space results, we will discuss its application to the torus.  

\subsection{Three dimensions}  \label{rg3}
For three-dimensional QFTs in $\mathbb R^{1,2}$, given a smooth entangling region $A$ characterized by a unique length scale $R$, 
the REE is defined as \cite{Liu2012,CH_F}   
\begin{equation} \label{Fr}
  \mathcal{F}(R)=-S(R)+R \frac{\partial S(R)}{\partial R}\, ,
\end{equation}
where $S(R)$ is the full EE associated with region $A$.
Crucially, $A$ is assumed to be ``scalable'' \cite{Liu2012}, meaning that its shape does not change as $R$ is varied.
This guarantees that we can meaningfully compare the entanglement associated with $A$ at different length scales. 
When applied to a disk-shaped region $A$ of radius $R$, cut out of the vacuum of a CFT \cite{CH_F,Liu2012}, $\mathcal F$ isolates the 
regulator- and $R$-independent term, \ie $\mathcal{F}(R)=F$.  
In fact, $\mathcal{F}(R)$ interpolates between $F_{\rm UV}$ and $F_{\rm IR}$ when evaluated along a RG flow linking the UV and IR CFTs. 
As proved in \cite{CH_F}, it does so in a monotonously decreasing manner, which establishes a ``c-theorem'' for general three-dimensional CFTs, called the F-theorem (in its strong form).  

When considering the spatial manifold to be a torus $\mathbb T^2$ instead of $\mathbb R^2$, and region $A$ to be a cylinder (\rfig{fig:torus3d}), 
we need to keep $\theta$ and the aspect ratio $b$ constant as $L_y$ (which plays the role of $R$) changes, 
in order to have a scalable region. 
In other words, we fix the dimensionless shape parameters $\{\theta,b\}$ and let $L_A = L_y\cdot \theta b/(2\pi)$ as well as $L_x=L_y\cdot b$. 
In analogy with $\mathcal F(R)$ \req{Fr}, we can then define a \emph{renormalized} torus EE:
\begin{align}
  \chi_r(L_y;\theta,b) = - S(L_y) +L_y \frac{\partial S(L_y)}{\partial L_y}\,,
\end{align}
where we have omitted the dependence of the full EE, $S$, on the \emph{constants} $\theta,b$. 
$\chi_r$ explicitly depends on the scale $L_y$ when the theory is away from a conformal fixed point.   
Since $\chi_r$ is dimensionless, $L_y$ will appear in combination with the coupling
constants that determine the RG flow. At a CFT fixed point, $\chi_r$ naturally reduces to the $L_y$-independent value encountered earlier, $\chi(\theta)$.   

In the limit in which the spatial torus becomes very thin, $L_y\to 0$, $\chi$ reduces to a constant, $\chi=2\gamma$, when evaluated at a 
CFT fixed point --- see section \ref{cyl3}.
This universal constant $\gamma$ can be more directly isolated
by considering an entangling region $A$ that is a half-infinite cylinder cut out of an infinite cylinder, \req{ees2}, which we will focus on.   
A definition of the REE for this semi-infinite cylindrical region will then be
\begin{equation} \label{reecyl}
  \gamma_r(L_y)=-S(L_y)+L_y \frac{\partial S(L_y)}{\partial L_y}\, .
\end{equation}
Notice that unlike for the torus function, there is no dependence on shape parameters.  
Further, just like for the EE of a disk, when \req{reecyl} is applied to the CFT expression \req{ees2},   
it isolates the universal term, \ie $\gamma_r (L_y)=\gamma$.  In fact, the previous definition can be generalized as follows:
\begin{align} \label{gy21}  
  \gamma_r^{(\alpha)} (L_y) = -S + L_y\frac{\partial S}{\partial L_y} + \alpha L_y^2 \frac{\partial^2 S}{\partial L_y^2} \,.
\end{align}
Compared with \req{reecyl}, $\gamma_r^{(\alpha)}$ contains a new second-derivative term parametrized by $\alpha$; it reduces to
the previous definition when $\alpha=0$. 
We see that the second order derivative term annihilates the area law contribution, meaning that it is cutoff independent
as $\epsilon\to 0$. We can view \req{gy21} as a natural extension of $\gamma_r$ in terms of a gradient expansion in $L_y$. 
One could in principle consider higher order terms, but we shall restrict our analysis to the present truncation.
$\gamma_r^{(\alpha)}$ is well-defined in the $\epsilon\to 0$ limit, and takes the value $\gamma$ at CFT fixed points. 
Further, just like $\gamma_r$, it is linear in $S$, implying that is is additive for decoupled theories. 
In section \ref{holorg} we will compute $  \gamma_r(L_y)$ for a $3d$ CFT deformed by a relevant scalar operator 
of scaling dimension $1/2<\Delta<3$. As we shall see, this quantity will be well-defined for all the allowed values 
of $\Delta$ and, in particular, independent of the cut-off $\epsilon$.  

\subsection{Four dimensions}
\label{rg4} 
Let us now explore the four-dimensional case. We are interested in the behavior of the EE under RG flows on $\mathbb{T}^3$.
In this case, the REE introduced  
in \cite{Liu2012} for four-dimensional QFTs in infinite Minkowski space, $\mathbb R^{1,3}$, given a 
scalable smooth entangling region $A$ with a characteristic length scale $R$, reads
\begin{equation} \label{Fr2}
  \mathcal{S}(R)=\frac{1}{2}\left[R^2\frac{\partial^2 S(R)}{\partial R^2}-R \frac{\partial S(R)}{\partial R} \right],
\end{equation} 
where $S(R)$ is the full cutoff-dependent EE associated with $A$. When computed in a CFT, $\mathcal{S}(R)$ isolates 
the corresponding universal contribution: the coefficient of $\log(R/\epsilon)$.  
As we mentioned before, the situation for the entanglement on $\mathbb{T}^3$ that we have considered (\rfig{fig:torus4d}) 
is quite different from the situation just described, because in that case the CFT universal term is constant
as a function of $\epsilon$, \req{tete}, and not logarithmic. 

Anticipating our discussion of renormalized entropies on tori,
we note that the above definition of $\mathcal{S}(R)$ is somewhat unique, in the sense that the only combination of the form $1/2(\alpha_1 R^2\partial^2_R S(R)+\alpha_2 R \partial_R S(R)+ \alpha_3 S(R))$ which isolates the universal term for a CFT is given by $\alpha_1=-\alpha_2=1$, $\alpha_3=0$, \ie the function defined in \req{Fr2}.

In the case of interest for us, in which the spatial manifold is $\mathbb{T}^3$ and the 
entangling region $A$ wraps two of its cycles (\rfig{fig:torus4d}), the role of $R$ in the definition of the REE 
can be played by one of the 4 length scales, $L_x,L_1, L_2$ or $L_A$. However, in order to have a ``scalable'' region (looking
the same for all sizes), we need to fix the 3 shape parameters $\{\theta,b_1,b_2 \}$,
leaving us with a single characteristic length scale.   
For example, choosing $L_1$ as our variable length, all the other lengths will scale linearly with $L_1$: $L_A=L_1 \cdot b_1\theta/(2\pi)$, $L_x=L_1 \cdot b_1$, $L_2=L_1\cdot b_1/b_2$. 
We can now define the following family of \emph{renormalized} torus functions as
\begin{align}\label{reecyli0}
\chi^{(\alpha)}_{r}(L_1;\theta,b_1,b_2)=\frac{1}{2}\Big[&\alpha L_1^2 
    \frac{\partial^2 S(L_1)}{\partial L_1^2}\\ \notag &\left.+(1-\alpha) L_1 \frac{\partial S(L_1)}{\partial L_1}-2 S(L_1)\right],
\end{align} 
where we omitted the dependence of $S$ on $\theta,b_1$ and $b_2$. We have introduced the free parameter $\alpha$. 
At a conformal fixed point, $\chi_r^{(\alpha)}$ reduces to $\chi(\theta)$ as given in \req{tete}.  
Note also that \req{reecyli0} is rather different from \req{Fr2}, as it always contains a term involving $S(L_1)$ while $\mathcal{S}(R)$ does not involve $S(R)$, but only its derivatives. Besides, there does not seem to be any reason to prefer any particular value of $\alpha$ in $\chi^{(\alpha)}_r$, as opposed to $\mathcal{S}(R)$ which is uniquely defined. Particularly simple cases correspond to $\alpha=0$ and $\alpha=1$, for which one finds 
\begin{align} \label{reecyli3}
  \chi^{(0)}_{r}(L_1) &=\frac{L_1}{2} \frac{\partial S(L_1)}{\partial L_1}- S(L_1)\, ,\\
  \chi^{(1)}_{r}(L_1) &=\frac{L_1^2}{2} \frac{\partial^2 S(L_1)}{\partial L_1^2}- S(L_1)\, .
\end{align}

Just like in three-dimensions, let us comment on the RG flow of a simpler quantity, obtained by considering
the thin torus limit: $L_{1,2}\to 0$, but with the ratio $L_2/L_1$ held fixed. In this limit, $\chi\to 2\gamma$,
where $\gamma$ can be obtained more simply by considering a half-infinite cylindrical region $A$, 
as discussed in section \ref{cylicio}. As opposed to the three-dimensional case, $\gamma$ depends in this case 
on a single shape parameter, namely $r=L_2/L_1$ --- see \req{gamba}. Hence, we define the renormalized $\gamma$ as
\begin{multline}\label{reecyli}
  \gamma^{(\alpha)}_{r}(L_1;r)=-S(L_1)\\ + (1-\alpha) \frac{L_1}{2} \frac{\partial S(L_1)}{\partial L_1} 
+\alpha \frac{L_1^2}{2} 
  \frac{\partial^2 S(L_1)}{\partial L_1^2}
\end{multline}
which of course reduces to $\gamma$ when applied to the CFT expression \req{ees33}. 
Indeed, at a CFT fixed point $\partial^n \gamma/\partial L_1^n=0$ for $n\geq 1$, since at that point
$\gamma$ only depends on the ratio $r=L_2/L_1$, which is held fixed.

\section{Holographic RG flow}\label{holorg}
We now analyze the RG flow of the universal EE associated with the bipartition of an infinite cylinder into two equal halves, $\gamma$, 
for a $3d$ CFT holographically dual to Einstein gravity in $4d$ \footnote{The interplay between RG flows and EE in the holographic context has been often considered in the past. See \eg \cite{Klebanov:2012yf,sinha10,sinha11,Myers:2012ed,Liu:2013una,Taylor:2016aoi,Casini:2015ffa} and references therein.}. In order to do so, we perturb  
our original holographic UV CFT by a relevant bosonic scalar operator $O(x)$, with scaling   
dimension $1/2\leq\Delta <3$ (the lower bound follows from unitarity).
This corresponds to the following deformation on the CFT side of the duality:    
\begin{align} \label{S_pert}
  \mathsf S=\mathsf S_{\rm CFT} + \lambda \int d^dx \mathcal\,  O(x)\,,
\end{align}
where $\lambda$ is the uniform coupling constant.  
According to the holographic dictionary \cite{Witten,Maldacena,Gubser}, $O(x)$ is dual to a mass-$m$ bulk scalar field with 
$m^2=\Delta(\Delta-3)$. Recall that we have
set $L_{\rm AdS}=1$. The bulk action now reads 
\begin{equation}\label{4dact}
 I=\int \frac{d^{4}x\sqrt{-g}}{16\pi G}\,\left[6+R-\frac{1}{2}\partial_{\mu}\phi \partial^{\mu}\phi
   -\frac{m^2}{2}\phi^2+\dotsb \right].  
\end{equation}
As a result, the scalar field $\phi$ will perturb the bulk AdS-soliton geometry; we parametrize the perturbed metric as
\begin{align}\label{solitin22}
ds^2 =\frac{1}{z^2}\left[\frac{dz^2}{f \cdot g_1(z)}+dx^2+ f\cdot g_2(z)\, dy^2-dt^2 \right], 
\end{align}
where again $f=1-z^3/z_h^3$ and $g_i(z)= 1+ \mathcal{O}(z^{\sigma_i})$ as $z\to 0$, with $\sigma_i>0$, 
\ie we require that AdS$_4$ is recovered (locally) on the boundary.  
Imposing regularity at $z=z_h$ leads to the relation 
\begin{equation} \label{lbb2} 
  z_h=\frac{3L_y}{4\pi}\sqrt{g_1(z_h)g_2(z_h)}\,. 
\end{equation} 

 The EE for a cylinder region cut out of the torus, \req{solitin22}, is given by
 \begin{equation}\label{sss}
S=\frac{L_y}{2Gz_h \xi^{1/3}}\int_{\epsilon/(z_h\xi^{1/3})}^{1}\frac{d\zeta \sqrt{(1-\xi \zeta^3)g_2(\zeta)}}{\zeta^2\sqrt{g_1(\zeta)\tilde P(\xi,\zeta)}}\, ,
\end{equation}
 where now $\tilde{P}=(1-\xi \zeta^3)g_2(\zeta)-(1-\xi)\zeta^4$, and the relation between $L_A$ and $\xi$ reads in this case
 \begin{equation}
\frac{L_A}{2} =\int_0^1 \frac{\xi^{1/3}z_h \sqrt{1-\xi}\ \zeta^2d\zeta}{\sqrt{g_1(\zeta)(1-\xi \zeta^3)}\sqrt{\tilde{P}(\xi,\zeta)}}\, .
\end{equation}
 These expressions reduce to \req{ees} with \req{casi} and \req{lali} respectively when $g_1=g_2=1$. 
 In the thin torus limit, $S$ will contain a constant term corresponding to $2\gamma$ that we wish to extract. 
This can be extracted from \req{sss} by setting $\xi=1$:
 \begin{align}
S(\xi=1)=\frac{L_y}{2Gz_h}\int_{\epsilon/z_h}^{1}\frac{d\zeta }{\zeta^2\sqrt{g_1(\zeta)} }\, .
 \end{align}
To make further progress in evaluating this integral, we expand in the coupling $\lambda$, \req{S_pert}. 
The dimensionless expansion parameter is $\lambda z_{h0}^{3-\Delta} \propto \lambda L_y^{3-\Delta}$,
since $L_y$ is the only remaining IR scale in the problem.
We define the location of the unperturbed ``horizon''
\begin{align} \label{zh0}
  z_{h0}=\frac{3L_y}{4\pi}\,,
\end{align}
to distinguish it from $z_h$. The latter will be altered from its bare value $z_{h0}$ due to the backreaction of the scalar on the metric.
Expanding in $\lambda z_{h0}^{3-\Delta}$, the EE becomes
\begin{align} \label{S-expanded}
 S = \frac{L_y}{2G\epsilon}-\frac{L_y}{2G} \left[\frac{1}{z_h}+
\int_{\epsilon/z_{h0}}^{1}\frac{d\zeta }{\zeta^2} \frac{h_1(\zeta)}{2z_{h0}}z_{h0}^{2(3-\Delta)}\lambda^2\right],
\end{align}
where we expanded the functions appearing in the metric  as
\begin{equation}\label{ghh}
g_i(z)=1+h_i(\zeta)\; z_{h0}^{2(3-\Delta)} \lambda ^2 \, , \quad i=1,2\, ,
\end{equation}
anticipating that the leading correction will be at order $\lambda^2$, a fact we shall soon confirm.
In \req{S-expanded} and it what follows, 
\begin{align}
 \zeta=z/z_{h0} 
\end{align}
is defined using the bare ``horizon'' value \req{zh0}. From \req{S-expanded}, we see that in order to determine the universal EE, 
we need to obtain 1) the $\lambda$-corrected value of $z_h$,
2) the full $\zeta$-dependence of $h_1(\zeta)$ in the bulk, $0\leq \zeta\leq 1$. 
The situation is thus more complicated than in the analogous holographic RG calculation for the disk EE,
where one only needs to know the near boundary behavior of the metric correction 
(a single one suffices in that case). These extra complications arise because of the additional 
length scale, $z_h$, arising from the compactification. 
  
From \req{lbb2}, the value of $z_h$ including corrections of  $\mathcal{O}\big(z_{h0}^{2(3-\Delta)} \lambda^2\big)$ is
\begin{equation}
  \frac{z_h}{z_{h0}} = 1 + \frac{h_1(1)+h_2(1)}{2}\, z_{h0}^{2(3-\Delta)}\lambda^2\, .
\end{equation}
In order to get the full $\zeta$-dependence of the $h_i(\zeta)$, we need to solve the equation of motion of the scalar field on the entire spacetime, not only near the boundary.  
The equation for $\phi$ reads
\begin{equation} \label{phi-eom}
  \square \phi=\Delta(\Delta-3)\phi\,.
\end{equation}
Assuming $\phi$ depends only on the holographic coordinate $z$, the general solution to this equation in the unperturbed soliton background --- \ie in \req{solitin22} with $g_1(z)=g_2(z)=1$ and $z_h=z_{h0}$ --- reads 
\begin{equation}
\phi= \tilde{\phi}(\zeta)\cdot  z_{h0}^{3-\Delta}\, \lambda \, ,
\end{equation}
where the dimensionless function $\tilde\phi(\zeta)$ is $\lambda$-independent, and given by ($\Delta\!\neq 3/2$)
\begin{align}\label{fgf2}\notag
\tilde{\phi}(\zeta) &= \zeta^{3-\Delta}\cdot \, _2\mathsf{F}_1\!\left[1-\tfrac{\Delta}{3},1-\tfrac{\Delta}{3};2 \left(1-\tfrac{\Delta}{3}\right);\zeta^3\right]\\ &  - a\,\zeta^{\Delta} \cdot  \, _2\mathsf{F}_1\left[\tfrac{\Delta}{3},\tfrac{\Delta}{3};\tfrac{2\Delta}{3};\zeta^3 \right].
\end{align}
When $\Delta\!=\!3/2$, the two terms become degenerate, and the solution contains another term (see \cite{Todd16} for example).  
$_2\mathsf{F}_1[z_1,z_2;z_3;z_4]$ is the ordinary hypergeometric function.     
$\phi(z)$ scales as $\lambda (z^{3-\Delta} + a z_{h0}^{3-2\Delta} z^{\Delta}+\dotsb)$ near the boundary, where we have used the holographic
dictionary to relate $\lambda$ to the parameters entering in the solution of the equation of motion, \req{phi-eom}.  
The constant $a$ is found by asking that $\phi$ be regular near $\zeta=1$,   
\begin{align} \label{a-coeff}
  a= \frac{\Gamma \left(2-\frac{2 \Delta }{3}\right) \Gamma \left(\frac{\Delta
   }{3}\right)^2 }{\Gamma \left(1-\frac{\Delta }{3}\right)^2
   \Gamma \left(\frac{2 \Delta }{3}\right)} \, ,
\end{align}
which is strictly positive in the range $1/2\leq \Delta <3$, and vanishes linearly approaching $\Delta\!=\! 3$. 
In order to determine the backreaction on the metric, we need to solve Einstein's equations in the presence of this profile for $\phi(\zeta)$. These read
\begin{align}\label{einst}
&\ \ \ \ G_{\mu\nu}-3g_{\mu\nu}=\frac{1}{2}\partial_{\mu}\phi \partial_{\nu}\phi \\ \notag &-\frac{1}{4}g_{\mu\nu}\left[g^{\rho\sigma}\partial_{\rho}\phi \partial_{\sigma} \phi+\Delta(\Delta-3)\phi^2 \right]\, .
\end{align}
There are three non-trivial equations in \req{einst} --- corresponding to the  $zz$, $tt$ and $yy$ components --- but it can be shown that only two of them are independent when \req{fgf2} holds. We therefore have two equations 
for the two unknown functions, $h_i(\zeta)$. 
After some simplifications, those equations can be written at order $\mathcal{O}(z_{h0}^{2(3-\Delta)}\lambda^2)$ as
\begin{align}\label{ode}
  h_1' -\frac{3h_1}{\zeta (1-\zeta^3)}  &=Q(\zeta)\, , \\
  h_2'-h_1'&=- \zeta \frac{(\tilde\phi')^2}{2}\, ,
\end{align}
where 
\begin{align}
Q(\zeta)=\frac \zeta 4 \left[(\tilde\phi')^2+ \frac{\Delta(\Delta-3)\tilde\phi^2}{\zeta^2(1-\zeta^3)} \right]\, , 
\end{align}
and where all functions depend on $\zeta$ and we used the notation $()'=\partial()/\partial \zeta$. The first equation is solved by
\begin{align}\labell{h1}
  h_1(\zeta) =\int_{1}^{\zeta}d\bar \zeta\, \frac{ 1-\bar{\zeta}^{-3}}{1-\zeta^{-3}}\,Q(\bar{\zeta})\, , 
\end{align}
 which can be used to write $h_2(\zeta)$ as
 \begin{align}\labell{h2}
  h_2(\zeta)=h_1(\zeta) -\frac{1}{2}\int_{0}^{\zeta} d\bar{\zeta}\, \bar{\zeta}\, \tilde\phi'(\bar \zeta)^2\, .
\end{align}
 Both in \req{h1} and \req{h2}, we have imposed the boundary conditions $h_1(0)=h_2(0)=0$, which is consistent since $\phi(0)\!=\! 0$. 
In Fig.~\ref{h123}, we plot these functions for a particular value of the operator dimension, $\Delta\!=\! 2.8$.
 
 
 \begin{figure}
\label{fig:h03}\includegraphics[scale=.45]{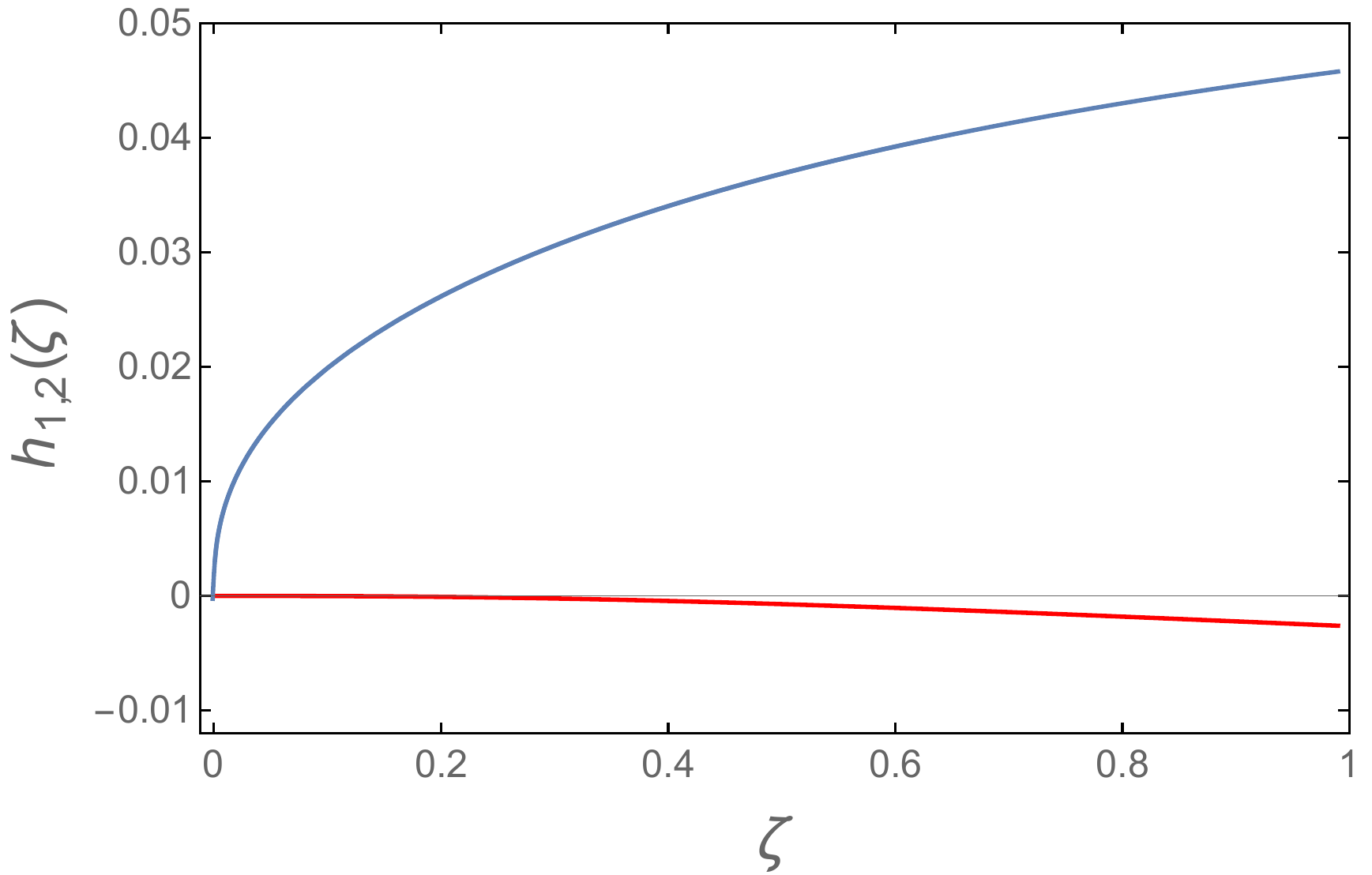}   
\caption{The metric perturbations $h_1(\zeta)$ (top, blue) and $h_2(\zeta)$ (bottom, red) for $\Delta=2.8$. Both vanish
at the boundary $\zeta\!=\! 0$.}      
\labell{h123} 
\end{figure}
 
From \req{h1} it is possible to obtain the small $\zeta$ behavior of $h_1(\zeta)$: 
\begin{align}\notag
h_1(\zeta)&=\frac{(3-\Delta)}{4} \zeta^{2(3-\Delta)}\left[  1+ \tfrac{9-2\Delta}{6} \zeta^3+\dotsb\right] 
\\ \notag &+\zeta^{2\Delta} \left[ D_{0\Delta}+D_{1\Delta}\zeta^3 +\dots\right]
\\ \labell{expo} &+\zeta^{3}\left[C_{0\Delta}+C_{1\Delta}\zeta^3+\dots\right] , 
\end{align}
where the dots mean $\mathcal{O}(u^{3k})$ terms with $k=2,3\dots$, and where 
\begin{align} \label{spurious}
  D_{0\Delta} = a^2 \Delta /4\,,
\end{align}
with $a$ defined in \req{a-coeff}.  
The other constants can also be explicitly given, but we refrain from doing so here.  
Since we need to evaluate $\int_{\epsilon/z_{h0}}^1 d\zeta\, h_1(\zeta)/\zeta^2$ in \req{S-expanded}, we observe from \req{expo} that the only terms 
susceptible of producing divergences in the $\epsilon\to 0$ limit (for $1/2\leq \Delta <3$) 
are those corresponding to $\zeta^{2(3-\Delta)}$ and $\zeta^{2\Delta}$. Indeed, we note that 
\begin{align}
  \int_{\epsilon/z_{h0}}^1 d\zeta \frac{\zeta^{2(3-\Delta)}}{\zeta^2}= \frac{1}{5-2\Delta}\left[1-\frac{z_{h0}^{2\Delta-5}}{\epsilon^{2\Delta-5}} \right]\, ,\\ \notag
  \int_{\epsilon/z_{h0}}^1 d\zeta \frac{\zeta^{2\Delta}}{\zeta^2}=\frac{1}{2\Delta-1}\left[1-\frac{z_{h0}^{1-2\Delta}}{\epsilon^{1-2\Delta}} \right]\, ,
\end{align} 
so the terms involving $\epsilon$ will diverge when $\epsilon\to 0$ for $\Delta\geq 5/2$ and $\Delta\leq 1/2$ respectively --- in the limiting cases, the divergences will be logarithmic instead. Taking this into account, we can rewrite \req{S-expanded} as
\begin{align}\label{eeed} 
S&=\frac{L_y}{2G\epsilon} 
-\frac{(3-\Delta)}{32(\Delta-5/2)}\, \frac{L_y\lambda^2}{\epsilon^{2\Delta-5}G}
\\ \notag &\quad +\frac{(\tfrac{3}{4\pi})^{6-4\Delta} D_{0\Delta}}{4(2\Delta-1)}\, \frac{L_y^{7-4\Delta}\lambda^2}{\epsilon^{1-2\Delta}G}
-2\gamma + \dotsb , 
\end{align}
where the dots denote terms subleading in $\epsilon$, and 
\begin{align}
\gamma=\frac{\pi}{3G}\left[1-\eta(\Delta)L_y^{2(3-\Delta)}\lambda ^2 \right]\, ,
\end{align}
and the $\Delta$-dependent constant $\eta(\Delta)$ reads
\begin{widetext}\begin{equation}\label{etaa}
\eta(\Delta)=\left( \frac{3}{4\pi}\right)^{\! 2(3-\Delta)}\! \left[\frac{h_1(1)+h_2(1)}{2}
+ \frac{3-\Delta}{16(\Delta-5/2)} - \frac{D_{0\Delta}}{2(2\Delta-1)}-\int_0^1\frac{d\zeta}{2\zeta^2}\left(h_1(\zeta)
-\frac{3-\Delta}{4}\zeta^{2(3-\Delta)}-D_{0\Delta}\zeta^{2\Delta} \right) \right].
\\ \end{equation}
\end{widetext}
 As we anticipated, \req{eeed} contains two new possible divergences $\sim \epsilon^{2\Delta-1}$ and $\sim \epsilon^{5-2\Delta}$. Let us start by emphasizing that neither violates the area law. The first divergence only arises 
when $\Delta \leq 1/2$, which corresponds to the unitarity bound for scalar operators. 
Therefore we shall work with $\Delta > 1/2$.
The second divergence is more important because it occurs in the allowed range, $5/2 \leq \Delta < 3$. 
We thus see that the naive subtraction of the area cannot yield a well-defined universal EE in this case. 
However, since this term scales linearly 
with $L_y$, any of the REE $\gamma_r^{(\alpha)}$ defined in \req{gy21} will make it disappear. 
Using the minimal prescription defined in \req{reecyl} --- corresponding to $\alpha=0$ in \req{gy21} --- we find the following renormalized EE
\begin{align}
\gamma_r(L_y) =\frac{\pi}{3G}\left[1-\eta_r(\Delta)L_y^{2(3-\Delta)}\lambda^2 \right]\, ,
\end{align}
where
\begin{equation}\label{renorG}
\eta_r(\Delta)=(2\Delta-5)\eta(\Delta)\, .
\end{equation}

\begin{figure}[h]
 \subfigure[]{\label{fig:h0}\includegraphics[scale=.48]{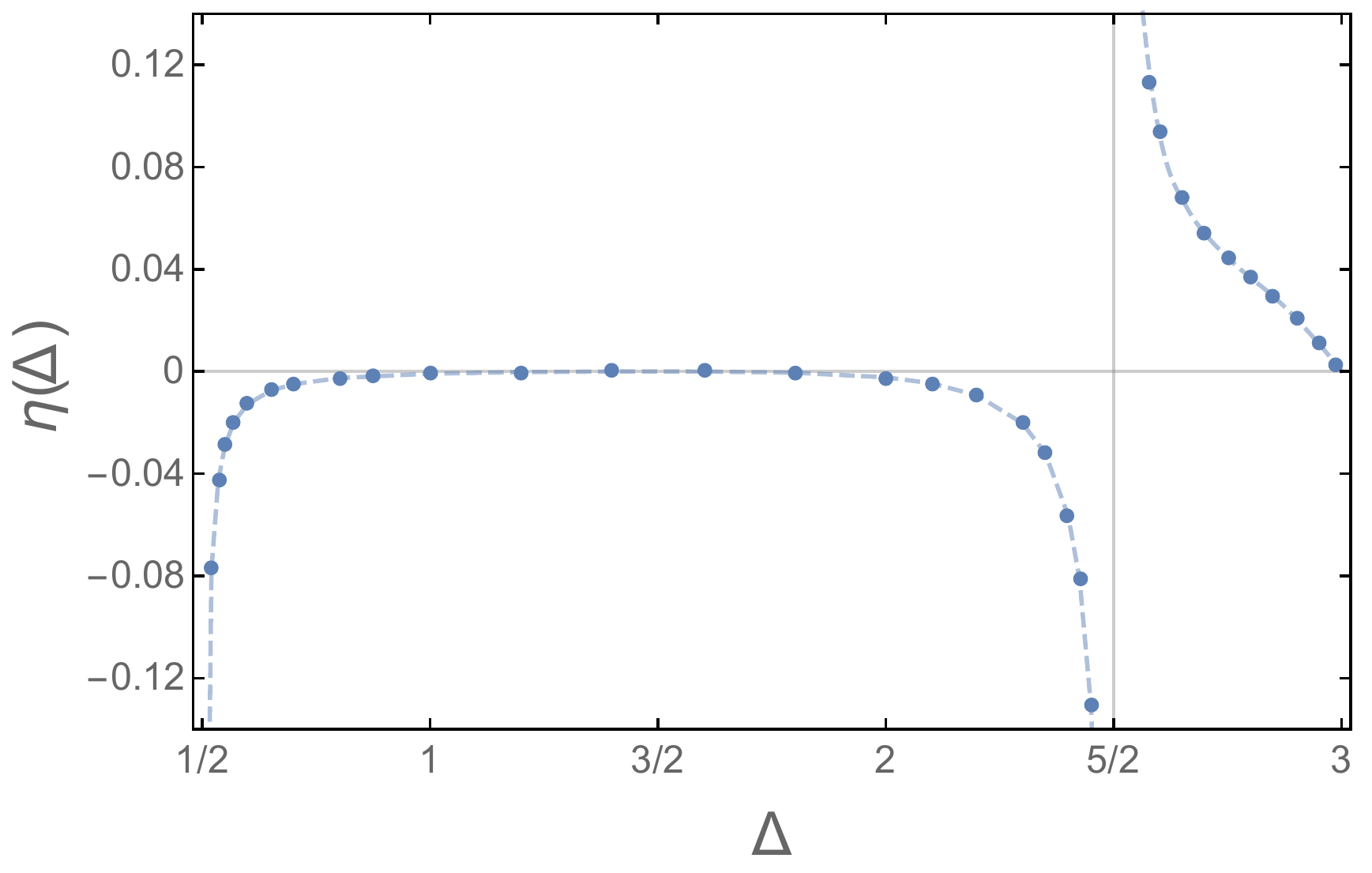}} 
 \subfigure[]{\label{fig:aE}  \includegraphics[scale=.48]{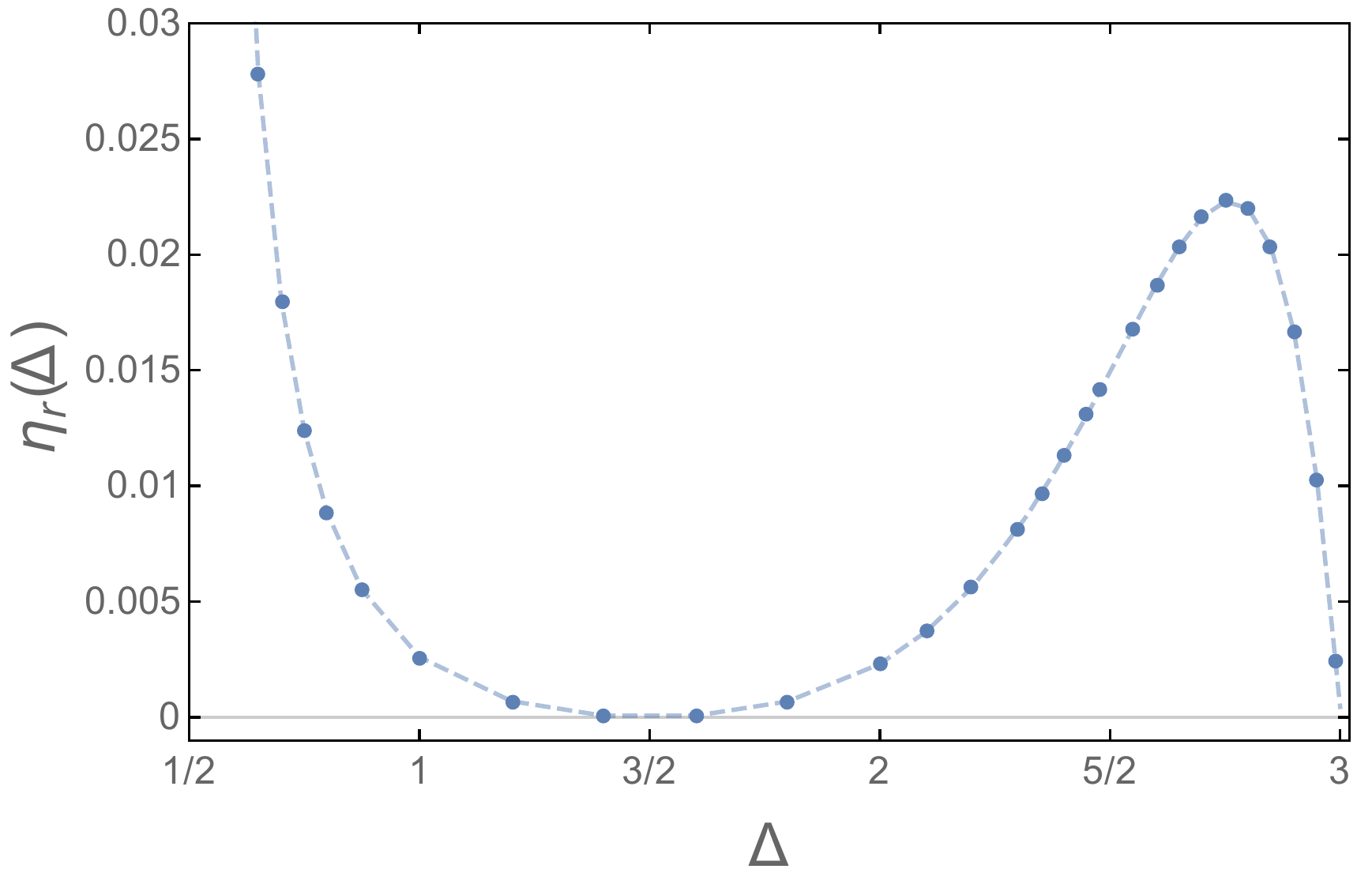}} 
\caption{{\bf Renormalized EE.} a)  We plot the bare $\eta(\Delta)$ as defined in \req{etaa} for the allowed range $1/2<\Delta<3$. At $\Delta=1/2^+$ and $\Delta=5/2^-$ the function diverges to $-\infty$ and conversely, it approaches $+\infty$ for $\Delta=5/2^+$.  b) We plot the renormalized version of the same quantity, $\eta_r(\Delta)$, as defined in \req{renorG}. The unphysical divergence at $\Delta=5/2$ is cured and the function is positive for all values of $\Delta$ except for $\Delta=3/2$ and $\Delta=3$, where it vanishes. Also, it diverges as $\Delta$ approaches the unitary bound $1/2$, as explained in the text. }  
\labell{gamma} 
\end{figure}

In Fig.~\ref{gamma} we compare $\eta(\Delta)$ with the renormalized version $\eta_r$. 
$\eta(\Delta)$ has an unphysical divergence at $\Delta=5/2$, which is cured in $\eta_r(\Delta)$. 
Note that such a spurious divergence appears as well in the non-renormalized universal term of the 
disk EE $F$ \cite{Liu2012,Klebanov:2012yf} --- see Appendix \ref{app1}. This is an indication that the naive subtraction of the divergent terms in the EE away from a conformal fixed point leads to unphysical results.
Further, the factor of $(2\Delta-5)$ allows us to to obtain the analytical result at $\Delta\!=\!5/2$:  
$\eta_r(5/2)\!=\! 3/(64\pi)\approx 0.0149$, in agreement with our numerical solution shown in \rfig{gamma}.

Interestingly, $\eta_r(\Delta)\geq 0$ for all dimensions in the 
range $1/2<\Delta<3$ --- it vanishes at $\Delta=3$, as expected for a marginal deformation, 
and at $\Delta=3/2$. Due to the positivity of $\eta_r$, the renormalized version of $\gamma$ decreases for all 
possible deformations with $1/2<\Delta<3$. In fact, the same decrease was recently obtained for the free scalar and Dirac fermion
CFTs under under mass deformations (and for all boundary conditions around the circle) \cite{W2}. 
It might be possible to prove this for general CFTs using methods of \cite{Faulkner14}. 
However, in going beyond the leading $\lambda$ correction to $\gamma_r$, the free CFTs exhibit an increase for certain, but not all,
boundary conditions. It would be interesting to verify whether this occurs in holography as well.  

It is also worth noting that $\eta_r$ diverges as one approaches  
the unitarity bound $\Delta=1/2$; such a divergence
was also found in the study of metric perturbations on the EE of spheres in holography \cite{Galante16}. 
It would be of interest to understand the physical mechanism behind this divergence at the unitary bound,
especially since not all quantities computed in such holographic theories will display it (see Ref.~\cite{Todd16} for example).
 Further, for this RG flow, $\gamma_r$ is \emph{stationary} at the UV fixed point: 
\begin{align}
  \left.\frac{\partial \gamma_r(t)}{\partial t}\right|_{t=0} = 0, \quad t\equiv \lambda L_y^{3-\Delta}\,,
\end{align}
The renormalized disk EE $\mathcal F$ is also stationary for the analogous holographic RG flow \cite{Nishioka_PRD_14}, 
when $3/2 < \Delta < 3$, 
see Appendix~\ref{app1}.   

\section{Entanglement estimate using the thermal entropy} \label{estimate}

We now use a simple approximation to get an estimate for the EE associated with a bipartition of the infinite cylinder
$\mathbb R\times \mathbb T^{d-2}$ into equal halves.  
Region $A$ is a half-infinite cylinder defined for $x\geq 0$, say. 
In this case, the Bisognano-Wichmann theorem \cite{BW} (here adapted from the infinite plane to the infinite cylinder) 
states that the reduced density matrix of $A$ is  
\begin{align} \label{BW}
  \rho_A =C\, \exp\left(-\int_{x>0}\! dx d^{d-2} y \, (2\pi x)\; \mathcal H(x, y) \right), 
\end{align}
where $\mathcal H = T_{00}$ is the energy density operator of the CFT on the cylinder $\mathbb R\times \mathbb T^{d-2}$;
$C$ is a constant. We can thus write the modular or entanglement Hamiltonian, defined via $\rho_A=C\exp(-H_E)$, 
as $H_E= \int_{x>0} dxd^{d-2}y\, \beta(x) \mathcal H(x,y)$, where
$1/\beta(x)=1/(2\pi x)$ is the $x$-dependent ``local temperature''. The following heuristic picture emerges:
degrees of freedom closer to the cut $x=0$ are at a higher ``local temperature'' and thus contribute more
to the EE compared to colder degrees freedom far from it. One can make the local picture more quantitative  
via the following ansatz for the EE, $S=-\tr(\rho_A\ln \rho_A)$:     
\begin{equation} \label{EE-therm}
  S = \int_{x>0} \! dxd^{d-2}y\;\, s_{\rm therm}\!\left( \frac{1}{2\pi x} \right),  
\end{equation}
where $s_{\rm therm}(T)$ is the thermal entropy density of the system at temperature $T$ on the \emph{infinite cylinder}. 
In \req{EE-therm}, the entropy is evaluated at a local temperature $T(x)=1/(2\pi x)$. This is in essence
a Thomas-Fermi approximate treatment of the exact $\rho_A$ given in \req{BW}:
after the trace over $\bar A$, one assumes that the system is locally in thermal equilibrium at temperature $1/(2\pi x)$. 
Such an ansatz has proven useful in various entanglement calculations. For instance, it yields 
the exact leading EE for the half-infinite interval in 1+1d CFTs \cite{Wong13}.      

We test these ideas using holographic CFTs. We first work in $d=3$,
where the thermal entropy on the infinite cylinder reads  --- see \eg \cite{Harlow:2013tf,Belin:2016yll},
\begin{align}
  s_{\rm therm} &= \frac{4\pi^2}{9G} T^2, &  &\!\!\!\textrm{if \quad $T>1/L_y$}\, , \label{s_therm} \\ 
  &= 0\,, & &\!\!\!\textrm{if \quad $T < 1/L_y$} \, .
\end{align}
It thus has an abrupt jump when $TL_y=1$, vanishing at small temperatures, where the AdS soliton
dominates the partition function.  
The large temperature value is naturally the entropy density of the finite-$T$ CFT in infinite 
flat space. By substituting \req{s_therm} into \req{EE-therm}, we obtain the EE:  
\begin{align}
  S &= L_y \int_\epsilon^{L_y/(2\pi)}\!\! dx\, \frac{4\pi^2}{9G} \left(\frac{1}{2\pi x}\right)^2\notag\\
  &= \frac{L_y}{9G\epsilon} - \frac{2\pi}{9G} \, .
\end{align} 
We thus recover the leading area law term. The subleading term is the universal EE, $\gamma_{\rm loc}$,
which has the same sign as in the exact holographic calculation, leading to
\begin{align}
  \gamma_{\rm loc} = \frac{2\pi}{9G} < \gamma_{\rm exact} = \frac{\pi}{3G} \, .
\end{align}
We thus find that the local entropy ansatz is lesser but close to the correct answer, 
$\gamma_{\rm loc}/\gamma_{exact}= 2/3$. 

In $d=4$, the thermal entropy on the infinite cylinder reads in turn --- see \eg \cite{Harlow:2013tf,Belin:2016yll},
\begin{align}
  s_{\rm therm} &= \frac{\pi^3}{4G} T^3, &  &\!\!\!\textrm{if \quad $T>1/L_1,1/L_2$}\, , \label{s_therm4} \\ 
  &= 0\,, & &\!\!\!\textrm{if \quad $T< 1/L_1$ or $T< 1/L_2$} \, .
\end{align}
In the above expressions, \req{s_therm4} is nothing but the thermal entropy of a black brane 
in AdS$_5$, 
which dominates the partition function over the corresponding soliton solutions for sufficiently
high temperatures. The thermal entropy vanishes instead when either of the two soliton geometries dominates. 
This difference in the thermal entropies of both solutions can be traced back to the fact that for 
the Euclidean black brane, regularity is achieved by imposing $z_h\sim T^{-1}$, while for the solitons one 
sets $z_h \sim L_i$ instead (where $L_i$ is the smallest dimension). This gives rise to a qualitatively different dependence on the temperature of the on-shell actions of both kinds of solutions --- and consequently, of the corresponding free energies and thermal entropies. Analogous comments apply in the $3d$ case discussed before, as well as for general higher dimensions, which we consider in Appendix~\ref{higherdim}.

Now, in the case $L_2>L_1$, using \req{s_therm4} we obtain from \req{EE-therm}
\begin{align}
  S = \frac{L_1 L_2}{16 G\epsilon^2} - \frac{\pi^2 L_2}{16L_1G} \, .
\end{align} 
The $L_1>L_2$ case is completely analogous: interchange $L_1\leftrightarrow L_2$ in the above expression.
Thus the universal term as obtained from the thermal Ansatz is again lesser than the exact result \req{gamba},
this time by a factor $1/2$, \ie $\gamma_{\rm loc}/\gamma_{exact}= 1/2$. We emphasize that 
that thermal entropy Ansatz yields exactly the same dependence on $L_{1,2}$.
In Appendix \ref{higherdim}, we show that the discrepancy is given by $\gamma_{\rm loc}/\gamma_{exact}= 2/d$ 
in general dimensions.
It would be interesting to see if the thermal entropy estimate 
yields a lower bound for $\gamma$ in other CFTs.



\section{Summary \& Outlook} \label{sec:disc}    

We have studied the universal contribution to the EE of 2+1$d$ and 3+1$d$ holographic CFTs on topologically 
non-trivial geometries, with a focus on tori.  More precisely, we have taken the spatial part of spacetime to be a torus 
$\mathbb{T}^{d-1}$, and have computed the EE for a bipartition of this space into two cylinders, $-\chi$. 
The cylindrical entangling region wraps $(d-2)$ of the non-contractible cycles. The geometries are illustrated in \rfig{fig:torus3d}
and \rfig{fig:torus4d} for the $d=3,4$ cases, respectively. 

We then studied the questions of how to characterize the RG flow of the torus EE $\chi$ in general QFTs (not only holographic ones).
We introduced a \emph{renormalized} EE in $3d$ and $4d$ that 1) is applicable to general QFTs, 2) resolves the failure of the area law subtraction away from fixed points, and 3) is inspired by the F-theorem. 
We have then employed this renormalized EE to study a simple RG flow for a $3d$ CFT with a holographic dual. 
The calculation was performed for a thin torus, in which case $\chi$ reduces to a geometry-independent constant $2\gamma$ at conformal fixed points.
The renormalized $\gamma$ was found to decrease monotonically for small deformations by a relevant operator (with any allowed dimension) 
away from the CFT fixed point. This is reminiscent of the results of Ref.~\cite{metlitski}, where $\gamma$ was compared between the Gaussian (free) and
the Wilson-Fisher CFT fixed point using the $\epsilon$-expansion. It was found that $|\gamma^{\rm WF}|<|\gamma^{\rm Gauss}|$,
where for some choice of boundary conditions the inequality holds without the absolute values. In line with this, 
recently it was shown \cite{WWS16}
that in the large-$n$ limit of the Wilson-Fisher fixed point, $\gamma^{\rm WF}= \mathcal O(n^0)$ while the UV Gaussian fixed point
has $\gamma$ scaling linearly with $n$.
However, in those works the value of $\gamma$ along 
the flow was not determined. As we saw, one is faced with the general question of how to define a suitable
renormalized EE away from fixed points. 
This question holds for essentially any entangling region, not only on the torus/cylinder. The prescription we have used
are sensible but a broader understanding is lacking, especially about the uniqueness of the renormalization procedure. 


Various concrete extensions of our work are possible. For one, it would also be interesting to extend our 
RG analysis to regimes beyond the thin-torus limit considered, and to $d=4$. 
More generally, extensions to other classes of deformations and to different holographic bulk theories in general $d$ --- \eg including higher-order gravity terms --- are conceivable. It would also be interesting to study the dependence of the
torus EE on the boundary conditions along the different cycles. In this respect, 
the twisted AdS soliton solutions \cite{Takabu} 
(obtained by a double-Wick rotation of a rotating black brane) could be useful.
Finally, one could study how the EE changes as the torus becomes non-rectangular, \ie sheared. \\

\begin{acknowledgments}     
We are thankful to A.~Bzowski, X.~Chen, T.~Faulkner, E.~Fradkin, D.~Galante, D.~Harlow, D.~Hofman, S.~Sachdev and S.~Whitsitt for many stimulating discussions.  
The work of PB was supported by a postdoctoral fellowship from the Fund for Scientific Research - Flanders (FWO). PB also acknowledges support from the Delta ITP Visitors Programme.
The work of WWK was funded by a Discovery Grant and a postdoctoral fellowship from NSERC, by MURI grant W911NF-14-1-0003 from ARO, 
and by a Canada Research Chair. 
The work of WWK was in part performed at the Aspen Center for Physics, which is supported by National Science Foundation grant PHY-1066293.
\end{acknowledgments}  

\onecolumngrid  
\appendix
\section{Disk EE in a perturbed holographic CFT} 
\label{app1}
We review the calculation of the renormalized EE of a disk region, $\mathcal{F}(R)$, 
in a three-dimensional holographic CFT deformed by a relevant scalar operator \cite{Liu2012,Klebanov:2012yf}.  
In doing so,
we shall emphasize the fact that the cutoff-independent term in the EE obtained by a naive subtraction of the area law, the ``naive $F$'', 
acquires an unphysical divergence 
for operators with scaling dimension $\Delta=5/2$, which disappears when considering $\mathcal{F}(R)$ instead. Other spurious 
cutoff-dependent divergences are also cancelled by $\mathcal F(R)$. 

The set-up is similar to the one in section \ref{holorg}, \ie we consider a CFT dual to the bulk theory \req{4dact}, where the mass-$m$ scalar field $\phi$ is dual to an operator $O(x)$ with scaling dimension $1/2\leq\Delta<3$ satisfying $\Delta(\Delta-3)=m^2$. 
The background spacetime is in this case pure AdS$_4$, and the perturbed metric can be parametrized in terms of a single function $g(z)$ as 
\begin{equation}\label{adspert}
  ds^2=\frac{L^2}{z^2}\left[\frac{dz^2}{g(z)}+dr^2+r^2d\theta^2-dt^2 \right],
\end{equation}
where $g(z)= 1+\mathcal{O}(z^{\sigma})$ as $z\to 0$, with $\sigma>0$. 
Now, assuming $\Delta>3/2$, the solution of the scalar equation of motion \req{phi-eom} in the unperturbed background --- namely, \req{adspert} with $g=1$ --- reads
\begin{equation}
 \phi(z) =  \lambda \cdot z^{3-\Delta} \, .
\end{equation}
Using this expression in Einstein's equations \req{einst}, one can find the backreaction on the metric, which is encoded in $g(z)$. The result is
\begin{align} 
  g(z) = \frac{ 1 + \Delta(3 - \Delta)z^{2(3-\Delta)}\lambda^2/12 }
{1- (3-\Delta)^2z^{2(3-\Delta)}\lambda^2 /12 } \,.  
\end{align} 
We will not expand in $\lambda$ at this point; we shall do so when we evaluate the EE. 
Using the Ryu-Takayanagi prescription \req{RyuTaka}, it is possible to show that the entanglement entropy of a disk region of radius $R$ is given by 
\begin{align}
S_{\rm disk}=\frac{\pi }{ 2G}\int_{\epsilon/R}^{1}d\zeta \frac{\sqrt{1-\zeta^2(1-g(\zeta))}}{\zeta^2 \sqrt{g(\zeta)}} \, ,
\end{align}
where we made the change of variables  
\begin{align}
   \zeta=z/R\,. 
\end{align}
The above integral can be evaluated in powers of the dimensionless constant $R^{2(3-\Delta)}\lambda^2$, which
we take to be small. We find $g= 1 +\zeta^{2(3-\Delta)}\lambda^2 R^{2(3-\Delta)}(3-\Delta)/4+
\mathcal O(\lambda^4 R^{4(3-\Delta)})$, 
and so the result for the EE at order $\mathcal{O}(R^{2(3-\Delta)}\lambda^2)$ reads
\begin{align}
  S_{\rm disk}\frac{2G}{\pi} = \frac{R}{\epsilon} - 1 - \frac{(3-\Delta)\lambda^2R^{2(3-\Delta)}}{8}\left[-\int_0^1 d\zeta \zeta^{2(3-\Delta)}
+ \int_{\epsilon/R}^1 d\zeta \zeta^{4-2\Delta} \right] + \dotsb, 
\end{align}
where the dots denote terms subleading in $\epsilon$ (at this order in $\lambda^2R^{2(3-\Delta)}$).
Simplifying, we find
\begin{equation} \label{S_disk_full}
  S_{\rm disk}=\frac{\pi R}{ 2G \epsilon}-\frac{ \pi(3-\Delta)}{32(\Delta-5/2)}\frac{R \lambda^2}{G\,\epsilon^{2\Delta-5}}-F +\dotsb,  
\end{equation}  
with the $\epsilon$-independent part:
\begin{align}
  F= \frac{\pi}{2G}\left[1-\frac{(3-\Delta)}{16(\Delta-5/2)(7/2-\Delta)}\lambda^2 R^{2(3-\Delta)} \right] . 
\end{align}
This $F$ contains the same suspicious divergence at $\Delta=5/2$ as was found for the non-renormalized $\gamma$ in section \ref{rg3}. 
In fact, for $\Delta<5/2$, this $F$ obtained by naively subtracting the area law would grow under RG flow, 
thereupon violating the F-theorem \cite{CH_F}. There is an even more important problem with the naive subtraction
of the $R/\epsilon$ area law: when $5/2 < \Delta < 3$, \req{S_disk_full} contains a term that 
diverges as $(R/\epsilon)^{2(\Delta-5/2)}$ \cite{Nishioka_PRD_14}, 
exactly like in the torus calculation, \req{eeed}.  
Both problems are resolved by considering the REE, $\mathcal{F}(R)= (R\tfrac{\partial}{\partial R} -1)S$ \cite{CH_F,Liu2012}. The renormalized answer becomes \cite{Klebanov:2012yf} 
\begin{equation}\label{fefi}
  \mathcal{F}(R)= \frac{\pi}{2G}\left[1- \frac{(3-\Delta)}{8(7/2-\Delta)}\lambda^2 R^{2(3-\Delta)}+ \dotsb\right],
\end{equation}
which contains a correction due to $\lambda$ that is negative, in agreement with the F-theorem \cite{CH_F}. We note that \req{fefi}
actually holds \cite{Faulkner14} for general CFTs in the case of a relevant deformation with $\Delta> d/2$, Eq.~\req{S_pert}.     

\section{Holographic torus entanglement in higher dimensions} \label{higherdim} 
In this appendix we extend the results of section \ref{holotorus} to an AdS$_{(d+1)}$ soliton spacetime, \ie we consider a $d$-dimensional holographic theory whose spatial dimensions form a $T^{d-1}$ torus and a cylindrical entangling region $A$ of dimensions $L_A\times L_1 \times L_2 \times \dots \times L_{d-2}$, \ie wrapping $d-2$ of the torus cycles.
The relevant solutions have the form
\begin{align}\label{solitii}
ds^2&=\frac{1}{z^2}\left[\frac{dz^2}{f}+g_{xx}dx^2+g_{ii}(d{y^i})^2-dt^2 \right]\, ,
\end{align}
with $i=1,\dots,(d-2)$ and where $f=1-(z/z_h)^d$ can appear in $g_{xx}$ or in one of the $g_{ii}$. In the former case, we require $L_x=4\pi/d\,z_h$. For the $(d-2)$ additional solutions, corresponding to writing $f$ in $g_{ii}$, we need to impose $L_{i}=4\pi/d\,z_h$ instead. The relevant solution is determined again by the minimal free energy condition. Analogously to the $d=3,4$ cases, when $L_x$ is the smallest in the set $\{L_x,L_{1},\dots,L_{{d-2}} \}$, we need to consider the solution with $g_{xx}=f$. The same applies completely analogously for the $L_i$. So in general we have $d-1$ different solutions which can be relevant depending on the case. 

Let us start assuming that $L_{x}$ is the smallest length. The metric is \req{solitii} with $g_{xx}=f$ and the rest equal to $1$.
The final result for the cylinder holographic EE is
\begin{equation}\label{sold3}
S=\left[\frac{ L_{1}\cdots L_{{d-2}}}{2(d-2)G }\frac{1}{\epsilon^{d-2}}-\chi(\theta) \right]\, ,
\end{equation}
where
\begin{equation}
\chi(\theta)=\frac{2^{2d-5}(\pi)^{d-2}  }{d^{d-2}b_1\cdots b_{d-2}(d-2) G \xi^{\frac{(d-2)}{d}}}\left[\int_0^1\frac{-(d-2)\,d\zeta}{\zeta^{d-1}}\left[ \frac{1}{\sqrt{1-\xi \zeta^d-(1-\xi)\zeta^{2(d-1)}}}-1\right]+1\right]\, ,
\end{equation}
with $\xi=(z_*/z_h)^d$ and where $b_i=L_x/L_{i}$ satisfies $b_i\leq 1$ for all $i$. The function $\chi(\theta)$ can be obtained using
\begin{equation}
\frac{2\pi L_A}{L_x}=\theta(\xi)=d\cdot \xi^{1/d}\sqrt{1-\xi}\int_0^1\frac{d\zeta \,\zeta^{d-1}\,(1-\xi \zeta^d)^{-1}}{\sqrt{1-\xi \zeta^d-(1-\xi)\zeta^{2(d-1)}}}\, .
\end{equation}
When the cylinder length is small, $\theta<<1$, it is possible to find a closed-form expression for $\chi(\theta)$. It reads
\begin{equation}\label{tt}
\chi(\theta)=\frac{1}{b_1\cdots b_{d-2} G} \left[\frac{(2\pi)^{d-2}\kappa_{(d-2)}G}{\theta^{d-2}}
+\sum_{k=1}a_k\, \theta^{d(k-1)+2}\right]\, ,
\end{equation}
where the $a_k$ are pure numbers, and $\kappa_{(d-2)}$ is the universal coefficient in the holographic EE of a multidimensional slab of dimensions $\ell\times L^{d-2}$, which reads \cite{RyuTaka2}
\begin{equation}
  S_{\rm strip}=-\kappa_{(d-2)}\frac{L^{d-2}}{\ell^{d-2}}\, ,\quad \text{where} \quad \kappa_{(d-2)}=\frac{2^{d-3}\pi^{\frac{d-1}{2}}\,\Gamma \left[\frac{d}{2 (d-1)}\right]^{d-1}}{(d-2)G\, \Gamma \left[\frac{1}{2 (d-1)}\right]^{d-1}}\, .
\end{equation}
 When $L_{1}$ is the smallest direction instead, (and analogously for any of the other $d-3$ extra directions different from $x$), we need to consider the geometry with $g_{11}=f$ and the rest equal to $1$.
The result for the entanglement entropy of the cylinder is again given by \req{sold3}, 
where now
\begin{align}
\chi(\theta)&=\tilde{\chi}(\theta)\, , \qquad 0<\frac{\theta}{2\pi}< \frac{p}{b_1} \, ,\\ \notag
\chi(\theta)&=\frac{2^{2d-5}\pi^{d-2}b_1^{d-3}}{G(d-2)d^{d-2} b_2\cdots b_{d-2}}\, , \qquad \frac{p}{b_1}<\frac{\theta }{2\pi}< \frac{1}{2} \, ,
\end{align}
and 
\begin{align}
\tilde{\chi}(\theta)&=\frac{ 2^{2d-5}\pi^{d-2} b_{{1}}^{d-3}}{G (d-2)d^{d-2} b_{2}\cdots b_{{d-2}} \xi^{\frac{(d-2)}{d}}}\left[\int_0^1\frac{-(d-2)d\zeta}{\zeta^{d-1}}\left[ \frac{\sqrt{1-\xi \zeta^d}}{\sqrt{1-\xi \zeta^d-(1-\xi)\zeta^{2(d-1)}}}-1\right]+1\right]\, ,\\
\frac{2\pi L_A}{L_x}&=\theta(\xi)=\frac{d\cdot  \xi^{1/d}}{b_1}\sqrt{1-\xi}\int_0^1\frac{d\zeta \,\zeta^{d-1}\,(1-\xi \zeta^d)^{-1/2}}{\sqrt{1-\xi \zeta^d-(1-\xi)\zeta^{2(d-1)}}}\, .
\end{align}
Again, $\chi$ for $\pi<\theta<2\pi$ can be obtained using the reflection property $\chi(2\pi-\theta)=\chi(\theta)$.
It can be checked that for $\theta\rightarrow 0$, the behavior of $\tilde{\chi}(\theta)$ reads
\begin{equation}\label{tt}
\tilde{\chi}(\theta)=\frac{1}{b_2\cdots b_{d-2}G} \left[\frac{(2\pi)^{d-2}\kappa_{(d-2)}G}{\theta^{d-2}b_1}
+\sum_{k=1}a_k\,b_1^{dk-1}\, \theta^{d(k-1)+2}\right]\, ,
\end{equation}
whose leading term is the same as that of $\chi(\theta)$ in \req{tt}, and where the $a_k$ are pure numbers. The value of $p$ is $\sim 0.2$ at least for the first higher-dimensional cases $d\geq4$.

In the cylinder limit, which corresponds to $L_x\gg L_i$ for all $i$, the result is also \req{sold3}, where now
\begin{align}
\chi(L_A/L_1)&=\tilde{\chi}(L_A/L_1)\, , \qquad 0<\frac{L_A}{ L_1}< p \, ,\\ \notag
\chi(L_A/L_1)&=2\gamma\, , \qquad p<\frac{L_A }{ L_1}< \infty \, ,
\end{align}
where
\begin{align}
\tilde{\chi}(L_A/L_1)&=\frac{ 2^{2d-5}\pi^{d-2}L_{2}\cdots L_{{d-2}} }{G (d-2)d^{d-2}  L_{{1}}^{d-3} \xi^{\frac{(d-2)}{d}}}\left[\int_0^1\frac{-(d-2)d\zeta}{\zeta^{d-1}}\left[ \frac{\sqrt{1-\xi \zeta^d}}{\sqrt{1-\xi \zeta^d-(1-\xi)\zeta^{2(d-1)}}}-1\right]+1\right]\, ,\\
\frac{2\pi L_A}{L_1}&=d\cdot  \xi^{1/d}\sqrt{1-\xi}\int_0^1\frac{d\zeta \,\zeta^{d-1}\,(1-\chi \zeta^d)^{-1/2}}{\sqrt{1-\xi \zeta^d-(1-\xi)\zeta^{2(d-1)}}}\, ,
\end{align}
and where
\begin{equation}
 \gamma=\frac{2^{2d-6}\pi^{d-2} L_2\cdots L_{d-2}}{(d-2)d^{d-2}G   L_1^{d-3}}
\end{equation}
is the constant contribution corresponding to the case in which the entangling region $A$ becomes a semi-infinite cylinder. 
Note again that we are considering here the case for which $L_1$ is the smallest dimension. Analogous expressions hold whenever any of the remaining $(d-3)$ dimensions is the smallest one.

Finally, let us point out that the thermal entropy estimate explained in section \ref{estimate} yields in general dimensions
\begin{equation}
 S=\frac{L_{1}\cdots L_{{d-2}}2^{d-3}}{d^{d-1}(d-2)G}\left[\frac{1}{\delta^{d-2}}-\frac{(2\pi)^{d-2}}{L_{1}^{d-2}} \right]\, ,
\end{equation}
where we used the thermal entropy expression on the infinite cylinder at sufficiently high temperatures, given by
\begin{equation}
 s=\frac{4^{d-2}\pi^{d-1}}{d^{d-1}G}T^{d-1}\, .
\end{equation}
We observe that the value of $\gamma$ obtained in this case reads
\begin{equation}
 \gamma_{\rm loc}=\frac{2^{2d-5}\pi^{d-2} L_2\cdots L_{d-2}}{(d-2)d^{d-1}G L_1^{d-3}}\, .
\end{equation}
We thus find that the local entropy ansatz gives an answer surprisingly similar to the correct one, namely $\gamma_{\rm loc}/\gamma_{exact}= 2/d$. 

\bibliography{quartic-refs}{}    
\end{document}